\documentclass{emulateapj}

\usepackage{graphicx}
\usepackage{amsmath}
\usepackage{natbib}
\usepackage{amssymb}

\usepackage[breaklinks,colorlinks,urlcolor=blue,citecolor=blue,linkcolor=blue]{hyperref}
\usepackage{hyperref}

\newcommand{\be}{\begin{eqnarray}}
\newcommand{\ee}{\end{eqnarray}}

\newcommand{\lp}{\left(}
\newcommand{\rp}{\right)}
\newcommand{\lb}{\left[}
\newcommand{\rb}{\right]}

\def\b{\boldsymbol}

\newcommand{\slugcom}{Submitted for publication in The Astrophysical Journal}
\slugcomment{\slugcom}

\begin{document}

\normalsize

% -----------------------------------------------------------
% -----------------------------------------------------------

\title{Implications from ASKAP Fast Radio Burst Statistics}

\author{Wenbin Lu\altaffilmark{1}}
\author{Anthony L. Piro\altaffilmark{2}}

\altaffiltext{1}{TAPIR, Walter Burke Institute for Theoretical Physics, Mail Code 350-17, Caltech, Pasadena, CA 91125, USA;\\
wenbinlu@caltech.edu}

\altaffiltext{2}{The Observatories of the Carnegie Institution for Science, 813 Santa Barbara St., Pasadena, CA 91101, USA;\\
piro@carnegiescience.edu}

% -----------------------------------------------------------
% -----------------------------------------------------------

\begin{abstract}

Although there has recently been tremendous progress in studies of fast radio bursts (FRBs), the nature of their progenitors remains a mystery.
% A fundamental question associated with this is the relation between FRBs that are observed to repeat and those that have only shown a single burst.
We study the fluence and dispersion measure (DM) distributions of the ASKAP sample to better understand their energetics and statistics. We first consider a simplified model of a power-law volumetric rate per unit isotropic energy $dN/dE\propto E^{-\gamma}$ with a maximum energy $E_{\rm max}$ in a uniform Euclidean Universe. This provides analytic insights for what can be learnt from these distributions. We find that the observed cumulative DM distribution scales as $N(>{\rm DM})\propto {\rm DM}^{5-2\gamma}$ (for $\gamma>1$) until a maximum $\rm DM_{\rm max}$ above which bursts near $E_{\rm max}$ fall below the fluence threshold of a given telescope. Comparing this model with the observed fluence and DM distributions, we find a reasonable fit for $\gamma \sim 1.7$ and $E_{\rm max}\sim 10^{33}\rm\, erg\, Hz^{-1}$. We then carry out a full Bayesian analysis based on a Schechter rate function with cosmological factor. We find roughly consistent  results with our analytical approach, although with large errors on the inferred parameters due to the small sample size. The power-law index and the maximum energy are constrained to be $\gamma\simeq 1.6\pm 0.3$ and $\mathrm{log}E_{\rm  max} \,\mathrm{[erg\,Hz^{-1}]} \simeq 34.1^{+1.1}_{-0.7}$ (68\% confidence), respectively. From the survey exposure time, we further infer a cumulative local volumetric rate of $\mathrm{log}N(E>10^{32}\rm \, erg\, Hz^{-1})\, \mathrm{[Gpc^{-3}\,yr^{-1}]}\simeq 2.6\pm 0.4$ (68\% confidence). The methods presented here will be useful for the much larger FRB samples expected in the near future to study their distributions, energetics, and rates.

% The similarity of this value to that of repeating FRBs argues that single bursts may share the same underlying physical mechanism with the repeaters. We also discuss the fluence distribution of the ASKAP sample, which allows us to estimate a maximum isotropic burst energy of $E_{\rm max}\simeq 2\times 10^{33}\,\rm erg\,Hz^{-1}$ as well as the volumetric rate of FRBs as a function of their energies. 
%\NEW{The errors of the inferred parameters are obtained from a Bayesian analysis. Our method can be applied to much larger FRB samples expected in the near future to study their distributions, energetics, and rates.}

\end{abstract}

	
% -----------------------------------------------------------
% -----------------------------------------------------------

\section{Introduction}
Fast radio bursts \citep[FRBs,][]{Lorimer07,Thornton13} are millisecond radio pulses with large dispersion measures (DMs) strongly suggesting an extragalactic origin. This has been directly confirmed by the repeater FRB 121102 \citep{Spitler16}, which has been localized to a $z=0.19$ galaxy \citep{Chatterjee17,Tendulkar17}. FRBs promise to provide amazing probes of the baryonic distribution across cosmological distances, but before their full potential can be reached, a fundamental understanding of their sources is required \citep[for an overview of the observations and potential progenitor models, see][]{Platts18, 2019A&ARv..27....4P}\footnote{\hyperlink{http://frbtheorycat.org}{http://frbtheorycat.org}}. A chief issue is the connection between repeating FRBs and those seemingly one-off bursts. Do all FRBs have the same source as the repeaters? What is the volumetric rate of FRBs as a function of their energies? And how do their nature and rate impact their utility as cosmological probes?

\begin{figure}
\includegraphics[width=0.48\textwidth,trim=0.0cm 0.0cm 0.0cm 0.0cm]{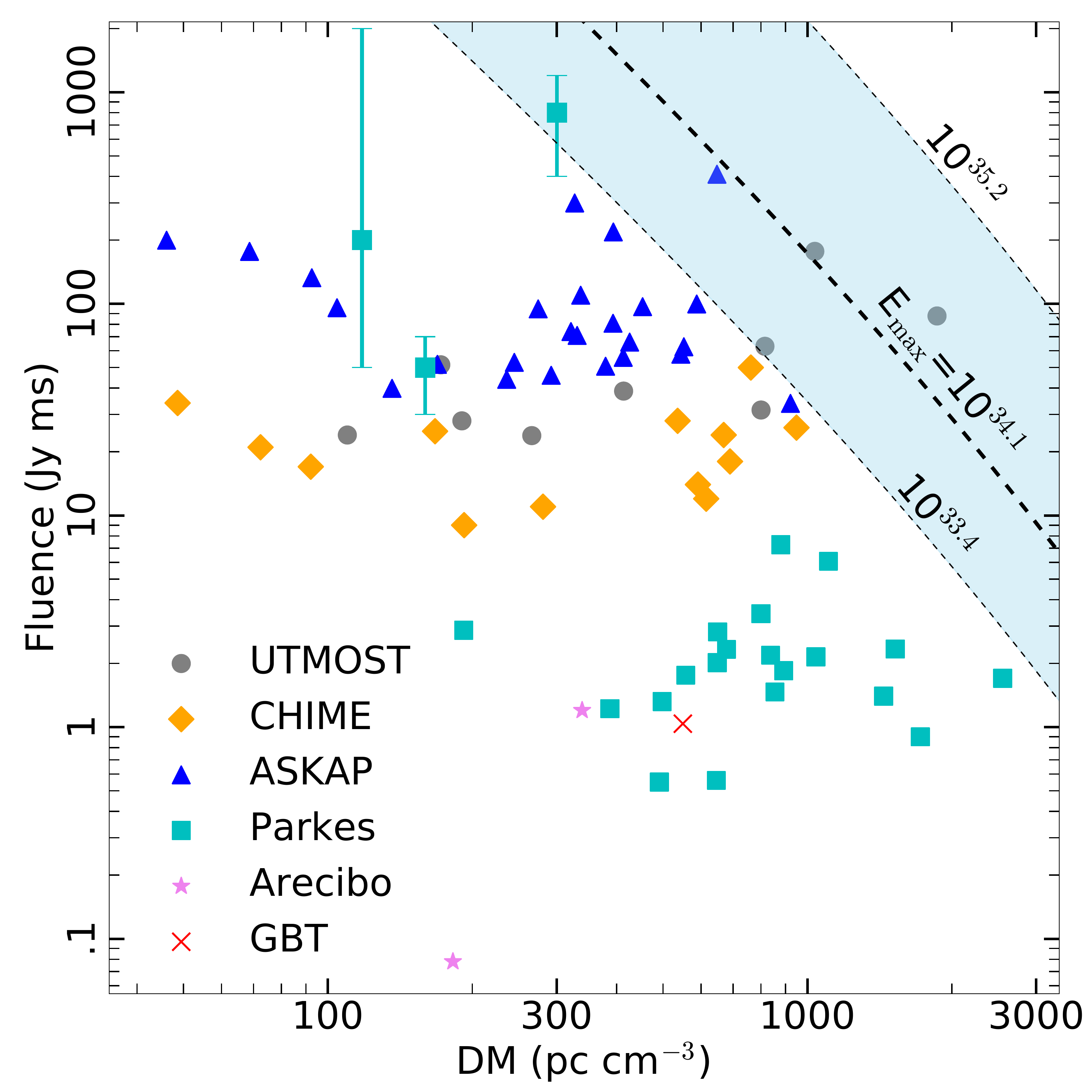}
\caption{The DM and fluence of the current sample of FRBs from the FRB Catalog \citep{Petroff16}. The three dashed lines show $F = E_{\rm max}/4\pi D_{\rm L}^2$ for $E_{\rm max} = 10^{33.4}$ ($-1\sigma$), $10^{34.1}$ (medium value), and $10^{35.2}\rm \,erg \, Hz^{-1}$ ($+1\sigma$) as constrained by our Bayesian analysis in \S4. The luminosity distance $D_{\rm L}(z)$ is based on the numerical relation $z(\rm DM)$ of \citet{Zhang18}. Three of the Parkes bursts had multi-beam detections and hence their fluences (shown with errorbars) had been corrected for antenna attenuation.}
\label{fig:dm_fluence}
\end{figure}

These questions have made discovering FRBs and measuring their physical properties some of the leading scientific goals of many current and future telescopes, such as Parkes \citep{Thornton13, Champion16, Bhandari18}, Arecibo \citep{Spitler16}, UTMOST \citep{Bailes17, Caleb17}, ASKAP \citep{Bannister17, Shannon18}, CHIME \citep{Amiri18}, FAST \citep{Li13}, and Apertif \citep{Maan17}. This has led to a rapidly growing but highly heterogeneous sample of FRBs. In Figure~\ref{fig:dm_fluence}, we summarize the DMs and fluences of currently published FRBs \citep{Petroff16}\footnote{\hyperlink{http://frbcat.org}{http://frbcat.org}}. The DMs are obtained by subtracting from the total measured values the contributions of the interstellar medium \citep{2002astro.ph..7156C} and halo ($\mathrm{DM}_{\rm halo}=30\rm \,pc\,cm^{-3}$) of the Milky Way.

Many studies have been carried out to statistically constrain the volumetric rate and luminosity/energy distribution function of the growing sample of FRBs, which may provide important clues about their progenitors \citep{Katz16, Lu16, 2017RAA....17....6L, Nicholl17, Macquart18, Luo18, 2019arXiv190210225B}. Such work has mainly focused on the Parkes sample. Although  Parkes  accounts for almost half of the bursts, there are many selection effects that make it fluence-incomplete \citep{Keane15, Patel18}, e.g., different backend instruments (AFB vs. BPSR) and galactic-latitude rate dependence \citep{Petroff14, Burke-Spolaor14}.
% Apparently, there is a deficit of bursts with low fluence and low DM. 
Likely the most serious issue is that the fluences of Parkes bursts are often lower limits (by assuming they occurred at the beam center) and are uncertain by up to an order of magnitude due to poor localization by single-beam detection. Three Parkes bursts (010724, 110214, 150807, shown in Figure~\ref{fig:dm_fluence} with errorbars) had multi-beam detections and their fluences included the correction of antenna attenuation \citep{Ravi16, Petroff19, Ravi19}.

Recently, the Commensal Real-time ASKAP Fast Transient (CRAFT) survey, targeting the brightest portion of the FRB population, has published 23 bursts --- the first and largest well-controlled sample to date. These bursts have well-measured fluences (uncertainty\footnote{One exception is FRB 170110 with a fluence of $200^{+500}_{-100}\rm\,Jy \, ms$ (90\% confidence). This large uncertainty is due to detection in a corner beam with poor localization \citep{Shannon18}. Nevertheless, it is well above the ASKAP fluence threshold and is hence included in our analysis.} of $\sim$20\%) because the overlapping beam arrangement allows the full focal plane to be uniformly sampled \citep{Shannon18}. The survey had high galactic latitude pointings $|b|=50\pm 5$~deg, which eliminates potential biases due to varying sky temperature and Galactic DM contribution. For the above reasons, we focus on the relatively uniform ASKAP sample \citep{Bannister17, Shannon18, 2019ApJ...872L..19M}. which can be shown to be roughly complete above a threshold fluence of $\sim$$50\,{\rm Jy\,ms}$.

This paper is organized as follows. We start with a simplified model to explain the observed DM distribution and the fluence distribution in \S2. Then, we discuss the inferred model parameters and implications in \S3. A numerical Bayesian analysis with a more sophisticated model is presented in \S4 to compare with our simpler analytic approach. 
Possible sample completeness issues are discussed in \S5. A few other potential caveats to keep in mind for future work are discussed in \S6. A summary is provided in \S7.

% There are two main approaches to understanding these mysterious sources: (1) construct  testable progenitor models from arguments based on energetics, timescale, radiation mechanism, and propagation effects; (2) use the observed statistical properties to infer the intrinsic luminosity/energy distribution function.

% The first approach has generated many models, such as coherent emission from the magnetosphere of neutron stars \citep[e.g.][]{Falcke14, Pen15, Cordes16, Dai16, Kumar17, 2017ApJ...836L..32Z}, emission from a relativistic (magnetar or AGN) outflow undergoing interaction with the surrounding medium \citep[e.g.][]{Lyubarsky14, Metzger17, Waxman17, Beloborodov17}. The rapidly growing data set, especially from the localized and well-monitored repeater FRB 121102 \citep{Spitler16, Chatterjee17, Tendulkar17, Michilli18}, has put increasingly stringent constraints on these models.

% \section{FRB Samples}

\section{DM and fluence Distribution}

In our sample, bursts with similar DM (or distance) show a large spread in fluences, which motivates us to consider a power-law volumetric rate of FRB events per unit (isotropic) energy
\be
	dN/dE = A E^{-\gamma},
\label{eq:simple_EDF}
\ee
with a maximum energy of $E_{\rm max}$ above which there are no FRBs.

\citet{Law17} found a power-law index of $\gamma\sim 1.7$ from the statistics of the first repeater FRB 121102 \citep[see also][]{2017JCAP...03..023W}. This is similar to the energy distribution of high-energy bursts from magnetars \citep{Turolla15}, which perhaps lends support for models which argue for magnetars as the progenitors of FRBs \citep{Popov10, Lyubarsky14, Kulkarni14, Pen15, Katz16, Kumar17, Metzger17, Beloborodov17}. Whether or not this is the case, such indices are natural consequences of self-organized critical phenomena \citep{Katz86, Bak87, Aschwanden16}. More recently, \citet{2019ApJ...877L..19G} reported 41 bursts from two hours of Arecibo observations\footnote{For comparison, earlier Arecibo searches at 1.4 GHz found 16 bursts in about 30 hours of observations \citep{2016Natur.531..202S, 2016ApJ...833..177S, 2017ApJ...846...80S}. }, most of which are faint ($E\sim 10^{29}\rm\, erg\,Hz^{-1}$), narrow-band ($\sim$$200\,$MHz), and low signal-to-noise ones found by careful visual selection against RFIs. The authors obtained a much steeper power-law index $\gamma\sim 2.8$ from this sample. It is possible that the inferred $\gamma$ is affected by the selection biases in different analyses and the non-Poissonian occurrence pattern \citep{2018MNRAS.475.5109O}. In this paper, we focus on the energy distribution of all FRBs (including repeaters and non-repeaters) and do not impose a prior on $\gamma$ based on the knowledge of FRB 121102.

Setting $F_{\rm th}$ as the threshold fluence for detection, the total number of events observed below a distance $D$ is
\be
\label{eq:NltD}
	&&N(<D) = \int_0^D 4\pi D^2 dD \int _{4\pi D^2 F_{\rm th}}^{E_{\rm max}} \frac{dN}{dE}dE
	\nonumber
	\\
	&&= {A E_{\rm max}^{1-\gamma}\over \gamma-1}\int_0^D 4\pi D^2 dD \left[\left(4\pi D^2 F_{\rm th} \over E_{\rm max}\right)^{1-\gamma} - 1\right].
\ee
Since bursts in our sample have low redshifts, we ignored the cosmological effects and changes of the rate normalization $A$ with distance (these will be included in the numerical analysis in \S4).

For sufficiently nearby bursts such that $4\pi D^2F_{\rm th}\ll E_{\rm max}$, the scaling of the cumulative number is
\begin{equation}
N(<D) \propto
    \begin{cases}
    D^3,\ \mathrm{for}\ \gamma < 1,\\
    D^{5-2\gamma},\ \mathrm{for}\ 1<\gamma<5/2,
    \end{cases}
\end{equation}
where the first case comes from the fact that most bursts are near $E_{\rm max}$ (like standard candles) and the second case reflects the effect of a wide distribution of burst energies.

For sufficiently large $D$, the rise of $N(<D)$ becomes shallower, because the maximum energy $E_{\rm max}$ limits the number of high-redshift bursts. The maximum distance a burst may be detected is given by
\be
\label{eq:Dmax}
	D_{\rm max} = (E_{\rm max}/4\pi F_{\rm th})^{1/2}.
\ee
The integral in Equation (\ref{eq:NltD}) gives
\be
	N(<D) &=& \frac{3N_{\rm tot}}{2(\gamma-1)} \lp\frac{D}{D_{\rm max}}\rp^3
	\nonumber
	\\
	&&\times\lb \lp \frac{D}{D_{\rm max}}\rp^{-2(\gamma-1)}
	- \frac{5-2\gamma}{3}  \rb,
	\label{eq:fit}
\ee
where the normalization $N_{\rm tot} = N(<D_{\rm max})$ is
\be
\label{eq:Ntot}
% 	N_{\rm tot} = \frac{2A E_{\rm max}^{1-\gamma}}{5-2\gamma}  \frac{4\pi D_{\rm max}^3}{3}.
	N_{\rm tot} = 2A E_{\rm max}^{1-\gamma}\,  4\pi D_{\rm max}^3/[3(5-2\gamma)].
	\label{eq:a}
\ee

% At this distance, the total number of FRBs is seen, so that $N_{\rm tot} = N(<D_{\rm max})$, allowing us to solve for the normalization of the energy distribution

% Then this integral can be solved
% \be
% 	N(<D) = \frac{(4\pi)^{2-\gamma}}{(5-2\gamma)(\gamma-1)} A F_{\rm th}^{-\gamma+1}D^{5-2\gamma}
% 		\nonumber
% 	\\
% 	- \frac{4\pi}{3(\gamma-1)} AE_{\rm max}^{-\gamma+1}D^3,
% 	\label{eq:n}
% \ee
% where this expression assumes $1<\gamma<5/2$. For sufficiently small $D$, the cumulative number scales as
% \be
% 	N(<D)\propto D^{5-2\gamma}.
% \ee
% In contrast, if there was just a fixed density of FRBs per unit volume, then $N(<D)\propto D^3$. The combination of a power-law energy distribution and fluence threshold for detection leads to a shallower slope for $N(<D)$. 

% For sufficiently large $D$, the rise of $N(<D)$ will become more shallow. As a check, we can solve for the maximum distance at which FRBs can be observed $D_{\rm max}$ by  taking the derivative of Equation (\ref{eq:n}) with respect to $D$ and setting it to zero, resulting in

% as expected. 
% Substituting this back into Equation (\ref{eq:n}) results in a simple fitting formula for the cumulative distribution with distance

\begin{figure}
\includegraphics[width=0.48\textwidth,trim=0.0cm 0.0cm 0.0cm 1.0cm]{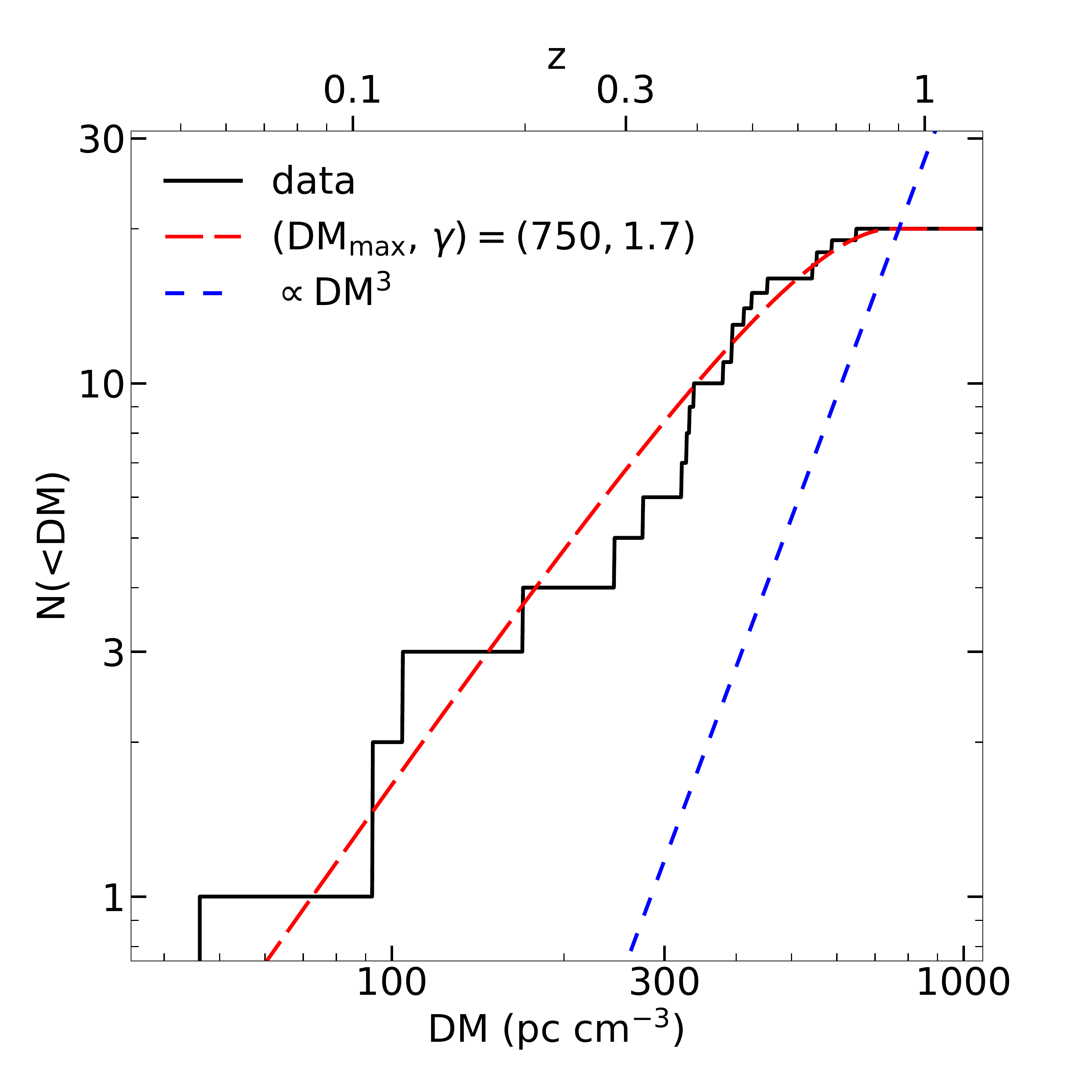}
\caption{Comparison of the ASKAP FRB sample (black solid line) with Equation (\ref{eq:fit}) (red dashed line) with $N_{\rm tot}=20$, ${\rm DM}_{\rm max} = 750{\rm\, pc\,cm^{-3}}$, and $\gamma=1.7$. We have subtracted the DM contributions from the Galactic interstellar medium \citep{2002astro.ph..7156C} and the Galactic halo ${\rm DM}_{\rm halo}=30\rm\,pc\,cm^{-3}$. Another case of $N(<D)\propto {\rm DM}^3$ is shown as a short-dashed blue line.
}
\label{fig:dm}
\end{figure}

Since at low redshifts we roughly have $D\propto z\propto {\rm DM}$, Equation~(\ref{eq:fit}) can easily be generalized as a function of $z$ or ${\rm DM}$. In Figure \ref{fig:dm}, we compare the expression in Equation (\ref{eq:fit}) to our ASKAP sample. We assume that the ASKAP sample is roughly complete above fluence $F_{\rm th}=50\rm\,Jy\,ms$ (see \S5), which encompasses 20 out of 23 bursts in the ASKAP sample (thus we take $N_{\rm tot}=20$). An energy distribution power-law index of $\gamma=1.7$ gives $N(<{\rm DM})\propto {\rm DM}^{1.6}$ in low-DM end.

In the future when the energy distributions of individual repeaters are better measured, we can compare the power-law index $\gamma$ obtained from the non-repeating sample with that of the repeaters. This provides a constraint on whether they belong to the same population.
% The similarity between observed distribution and Equation (\ref{eq:fit}) with $\gamma=1.7$ argues that FRB sources universally share the same energy distribution as the repeater FRB 121102 even if they may only generate single bursts \citep[see][for constraints on the repeating rate]{James19, Caleb19}. This may further suggest that the underlying mechanism is similar for both the repeaters and non-repeaters.
We also include a comparison to $N(<{\rm DM})\propto {\rm DM}^{3}$ (short-dashed blue line), which is appropriate if most bursts have characteristic energy. This is inconsistent with the ASKAP distribution, as mentioned by \citet{Li19}.

% \section{Fluence Distribution}

The total number of events above fluence $F$ is
\be
\label{eq:NgtF}
	N(>F) = \int_0^{\sqrt{E_{\rm max}\over 4\pi F}} 4\pi D^2 dD \int _{4\pi D^2F}^{E_{\rm max}} \frac{dN}{dE}dE.
\ee
Evaluating this integral and substituting the expression for $D_{\rm max}$ from Equation (\ref{eq:Dmax}) and $A$ from Equation (\ref{eq:a}), we obtain
\be
	N(>F) = N_{\rm tot} \lp F/F_{\rm th} \rp^{-3/2}.
	\label{eq:fluence}
\ee
Thus we expect the fluence distribution to be insensitive to $\gamma$ (assuming $\gamma<5/2$) and basically given by the value expected from 
a characteristic burst energy and Euclidean space. This is because, for Euclidean space, the number of bursts above a given fluence $F$ is dominated by those at the distance $\sim$$\sqrt{E_{\rm max}/4\pi F}$ where the brightest bursts are detectable \citep{Macquart18}.

\begin{figure}
\includegraphics[width=0.48\textwidth,trim=0.0cm 0.0cm 0.0cm 1.0cm]{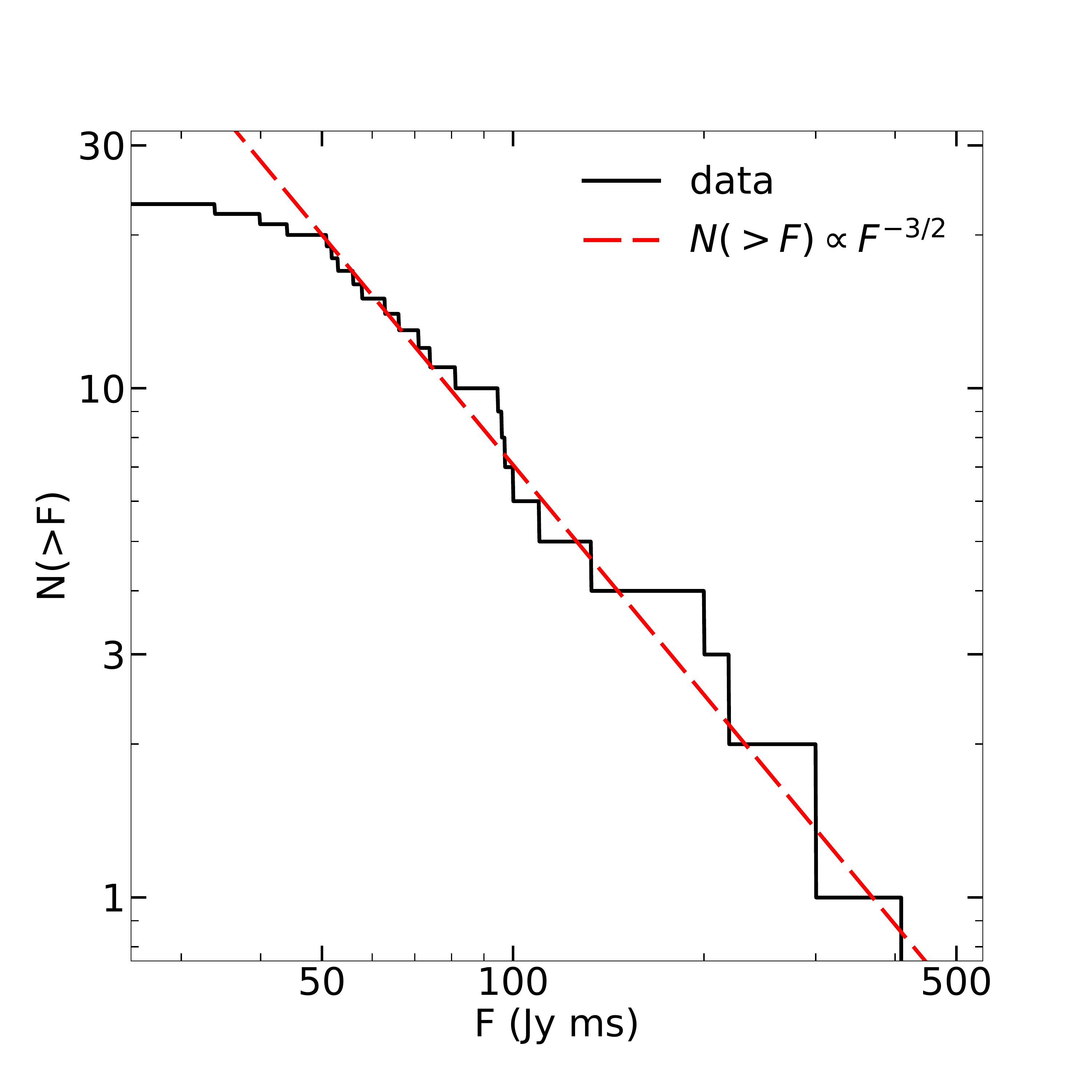}
\caption{Comparison of the ASKAP FRB sample (solid black line) with Equation (\ref{eq:fluence}) (dashed red line) using $N_{\rm tot}=20$ and $F_{\rm th}=50\,{\rm Jy\,ms}$. The sample is incomplete below the sensitivity threshold $F_{\rm th}$. }
\label{fig:fluence}
\end{figure}

In Figure \ref{fig:fluence}, we compare our fluence distribution given by Equation (\ref{eq:fluence}) with the ASKAP sample. A model with $N_{\rm tot}=20$ and $F_{\rm th}=50\,{\rm Jy\,ms}$ provides a good fit. The sample is incomplete below the sensitivity threshold $F_{\rm th}$ (see \S 5).

\section{Constraints on the Energy Distribution}

Combining $N_{\rm tot}$ with ${\rm DM}_{\rm max}$ and $F_{\rm th}$ estimated in the previous discussions allows us to constrain the energy distribution function of the FRB population. To relate luminosity distance and DM we use
\be
\label{eq:distance}
	D \simeq 6.7\mathrm{\,Gpc}\, \lp{\rm DM}/855\,{\rm pc\,cm^{-3}}\rp,
\ee
which is scaled to match the DM given by \citet{Zhang18} based on the latest Planck $\Lambda$CDM cosmology \citep{Planck16}. From the comparisons given in Figures \ref{fig:dm} and \ref{fig:fluence}, we can estimate the maximum isotropic FRB energy by using Equation~(\ref{eq:Dmax}),
\be
\label{eq:Emax}
	E_{\rm max} \simeq 2\times10^{33}{\rm\, erg\,Hz^{-1}}\frac{F_{\rm th}}{50\,{\rm Jy\,ms}}
	\lp\frac{{\rm DM_{\rm max}}}{750\,{\rm pc\,cm^{-3}}}\rp^2.
\ee
% As a greater sample of FRBs is collected, and with a better understanding of the ${\rm DM}_{\rm max}$ within which the sample is complete and the threshold fluence of a given observatory, then $E_{\rm max}$ can be better measured.
This in turn should be an important constraint on any emission model for FRBs \citep[e.g.,][]{lu18}.
% In Figure \ref{fig:dm_fluence}, we plot $F=E_{\rm max}/4\pi D^2$ (dashed line) using the $E_{\rm max}$ found here. This shows that the data strongly supports this rough estimate of $E_{\rm max}$ with the exception of one outlier that we discuss in Section 5.

On the other hand, the total exposure of the ASKAP survey was $\Omega T \simeq 5.1\times10^5\rm \,deg^2\,hr$ \citep{Shannon18},
which corresponds to an effective all-sky survey time $T_{\rm eff} \simeq 12.4\,\rm hr$.
% which can be converted to an all-sky rate of $38.8 \rm\,day^{-1}$ above the fluence threshold $F_{\rm th}\simeq 50\rm\,Jy\,ms$.
Since most bursts in our sample have moderate redshifts $z\sim 0.5$, we replace the Euclidean volume $4\pi D_{\rm max}^3/3$ in Equation (\ref{eq:Ntot}) with the comoving volume given by the Planck cosmology. For our fiducial values $N_{\rm tot}=20$, ${\rm DM}_{\rm max}\simeq 750\rm\,pc \,cm^{-3}$, and $\gamma \simeq 1.7$, the volumetric rate normalization is estimated to be
\be
    AE_{\rm max}^{1-\gamma}\simeq 90{\rm\,Gpc^{-3}\,yr^{-1}}.
\ee
Taking $E_{\rm max}\simeq 2\times10^{33}\rm \,erg\,Hz^{-1}$, we can further estimate the volumetric rate of FRBs above energy $E = 10^{32}E_{32}\rm\,erg\,Hz^{-1}$,
\be
N(>E)\simeq \frac{AE^{1-\gamma}}{\gamma-1}\simeq 1.1\times10^3 E_{32}^{-0.7} {\rm\,Gpc^{-3}\,yr^{-1}}.
\ee
This rate density is useful for comparison against potential FRB progenitors, although a critical unknown is how often each source may repeat at a given energy.
% For a typical spectral width of $\Delta \nu = 1$~GHz, we see that the rate of FRBs with isotropic energy $>10^{39}\rm\,erg$ is roughly $10\%$ of the core-collapse rate \citep{Li11, Kulkarni14} in the local Universe.
Non-repeating models based on rare events, such as long gamma-ray bursts or binary neutron star mergers \citep{Totani13, Zhang14, Wang16}, are inconsistent with the high rate of low-energy FRBs with $E\lesssim 10^{31}\rm\, erg\, Hz^{-1}$.

\begin{figure*}
\centering
\includegraphics[width=0.8\textwidth]{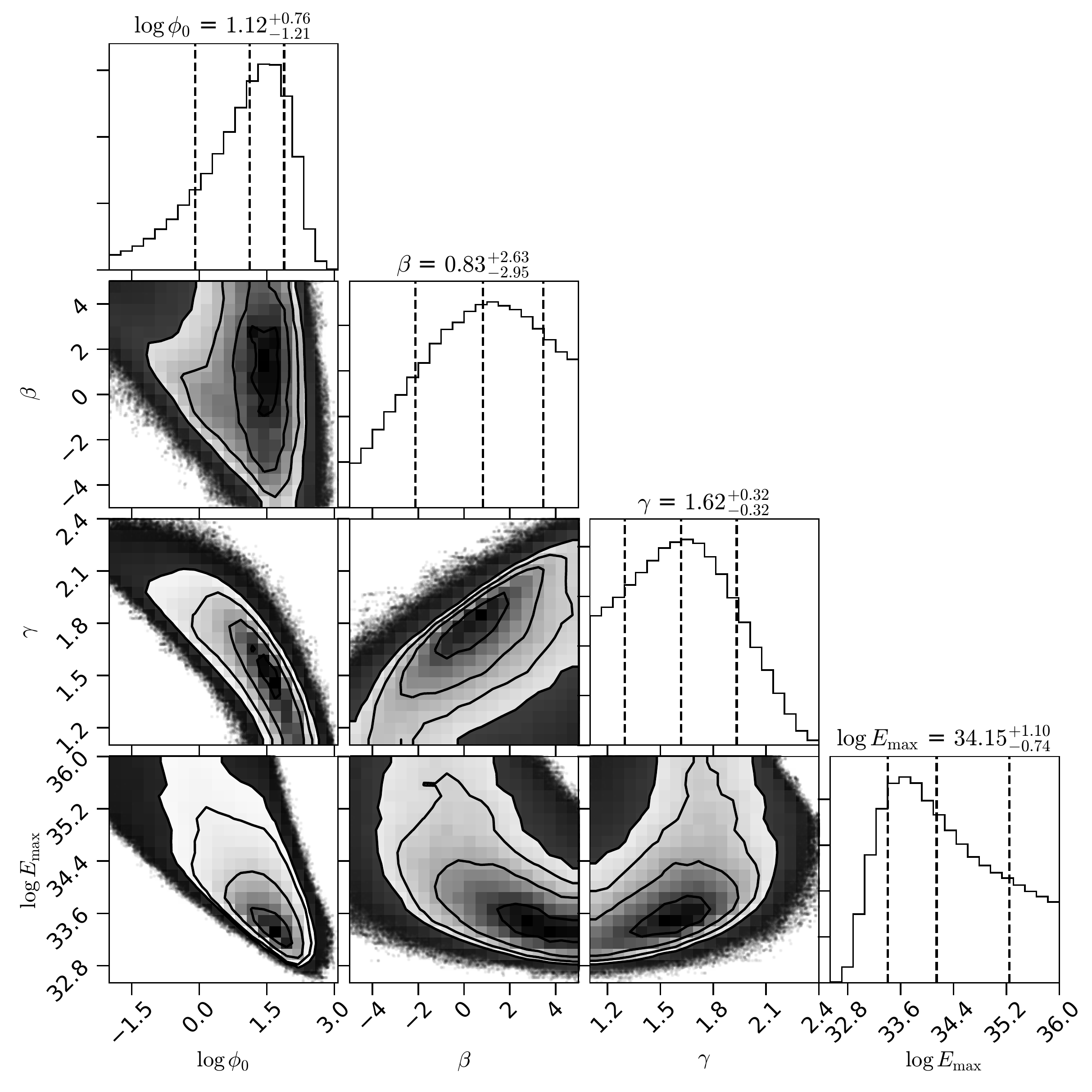}
\caption{MCMC sampling of the PDF (Equation \ref{eq:3}) of the parameters $\b{p}=(\mathrm{log} \phi_0$, $\beta$, $\gamma$, $\mathrm{log} E_{\rm  max})$, as constrained by the ASKAP FRB sample. The three vertical dashed lines in the marginal distributions marks where the cumulative density function (CDF) equals to $0.16,\ 0.5,\ 0.84$ (from left to right). The titles show the median (CDF$=0.5$) and the error range at 68\% ($1\sigma$) confidence level. This plot was generated with the public code \textit{corner.py} by \citet{corner}.}
\label{fig:corner}
\end{figure*}

\section{Bayesian Analysis of Full Parameter Space}

With a rough analytical understanding of the ASKAP FRB statistics in hand, we next consider a Schechter-like model
\begin{equation}
  \label{eq:1}
  {dN\over dE} = {\phi_0 (1+z)^{\beta}\over E_{\rm max}} \left(E\over E_{\rm max}\right)^{-\gamma} \mathrm{exp}\left(-E\over E_{\rm max}\right),
\end{equation}
where $\phi_0$ (in $\rm Gpc^{-3}\,yr^{-1}$) is the volumetric rate normalization at redshift $z=0$, $(1+z)^\beta$ represents possible evolution of FRB rate with redshift, $\gamma$ is the energy power-law index, and $E_{\rm max}$ is the maximum energy beyond which the rate cuts off exponentially. Hereafter, we adopt the latest Planck cosmology and the numerical integral relation $z(\mathrm{DM})$ in \citet{Zhang18}.

Equation~(\ref{eq:1}) predicts the distribution of redshifts (or DMs) and fluences for the observed bursts within the effective all-sky exposure time $T_{\rm eff}$,
\begin{equation}
  \label{eq:2}
  N(\leq z, \geq F) = \phi_0 T_{\rm eff} \int_0^zd z {dV\over dz} (1+z)^{\beta-1} \int_{x_{\rm min}}^{\infty} x^{-\gamma} \mathrm{e}^{-x} d x,
\end{equation}
where $dV/dz$ is the differential comoving volume, $x= E/E_{\rm max}$, and $x_{\rm min} = 4\pi D_{\rm L}^2(z)F(1+z)^{0.5}/E_{\rm max}$ with $D_{\rm L}(z)$ as the luminosity distance. Note that the $(1+z)^{0.5}$ factor in $x_{\rm min}$ is the k-correction assuming an average intrinsic spectrum\footnote{For $E_\nu=E_{\nu_0} (\nu/\nu_0)^{q}$ with arbitrary $q$ (where $\nu_0$ is the observer's reference frequency), we have $x_{\rm min} = 4\pi D_{\rm L}^2(z)F_{\nu_0}(1+z)^{-q-1}/E_{\nu_0,\rm max}$. The cosmological rate evolution parameter $\beta$ is degenerate with the spectral index $q$ due to k-correction. Variation in spectral index $\Delta q$ roughly corresponds to a linear shift in $\beta$ by $\Delta \beta \simeq (1-\gamma)\Delta q$. 
We note that the average FRB spectrum across a wide frequency range has not been well measured so far, so the cosmological evolution of FRB rate is not meaningfully constrained. The other three parameters are not affected by this degeneracy.} of $E_\nu\propto \nu^{-1.5}$ \citep{2019ApJ...872L..19M}.

We carry out a Bayesian analysis of the 4-dimensional parameter space $\b{p}=(\mathrm{log} \phi_0$, $\beta$, $\gamma$, $\mathrm{log} E_{\rm  max})$ by comparing the model prediction with the observed distribution as seen in Figure \ref{fig:dm_fluence}. The probability density function (PDF) of the parameter vector is given by (besides a normalization factor)
\begin{equation}
  \label{eq:3}
  f(\b{p})\propto L(\b{D}|\b{p}) f_0(\b{p}),
\end{equation}
where $L(\b{D}|\b{p})$ is the likelihood for the data $\b{D}$ to occur under a given parameter vector $\b{p}$, and we take a flat prior $f_0(\b{p})$ for sufficiently wide ranges of parameters: $\mathrm{log} \phi_0\, \mathrm{[Gpc^{-3}\,yr^{-1}]}\in (-2,\ 3.5)$, $\beta\in (-5,\ 5)$, $\gamma\in(1.1,\ 2.4)$, $\mathrm{log} E_{\rm  max} \,\mathrm{[erg\, Hz^{-1}]}\in (31,\ 36)$. The likelihood function measures the ``goodness of fit'' of how well the model at $\b{p}$ fits the data $\b{D}$. Since the data is a 2-dimensional distribution $N_{\rm obs}(\leq z, \geq F)$, information will be lost if we were to perform two marginalized 1-dimensional KS (or $\chi^2$) tests (as shown in Figures \ref{fig:dm} and \ref{fig:fluence}). In the spirit of the 2-dimensional KS test \citep{1983MNRAS.202..615P, 1987MNRAS.225..155F}, we take the following likelihood function
\begin{equation}
  \label{eq:4}
  L(\b{D}|\b{p}) = \min_{1\leq i\leq N_{\rm tot}}\left[P(k_1^{(i)},\ \lambda_1^{(i)}), P(k_2^{(i)},\ \lambda_2^{(i)})
\right],
% \vspace{0.15cm}
\end{equation}
where $\lambda_1^{(i)} = N(> z^{(i)}, \geq F^{(i)})$ and $\lambda_2^{(i)} = N(\leq z^{(i)}, \geq F^{(i)})$ are the expected number of detections in the first and second quadrants, $k_1^{(i)} = N_{\rm obs}(> z^{(i)}, \geq F^{(i)})$ and  $k_2^{(i)} = N_{\rm obs}(\leq z^{(i)}, \geq F^{(i)})$ are the actual number of detections, $i$ is the index of the observed bursts in our sample, and $P(k, \lambda) =\lambda^k \mathrm{e}^{-\lambda}/k!$ is the Poisson PDF. The cumulative numbers of observables ($z$ or $F$) are not independent of each other \citep{1967Natur.216..877J} and are hence only used once in the likelihood function in Equation (\ref{eq:4}).

The number of detections $N(> z^{(i)}, \geq F^{(i)})$ is calculated by integrating Equation (\ref{eq:2}) from the redshift of the $i$-th burst $z^{(i)}$ to a maximum value of $z_{\rm max}=2$. The precise value of $z_{\rm max}$ is not important, as long as it is much larger than the maximum redshift $\max(z^{(i)})\sim 0.7$ in our sample. We do not make use of the cumulative numbers in the third and fourth quadrants $N(\leq z^{(i)}, < F^{(i)})$ and $N(> z^{(i)}, < F^{(i)})$, because these numbers are subjected to uncertainties of the telescope threshold fluence $F_{\rm th}$ (see \S5 for a discussion).

We then generate $10^6$ Markov-Chain Monte Carlo (MCMC) samples of the parameter vector $\b{p}$ based on the PDF $f(\b{p})$ in Equation (\ref{eq:3}), as shown in Figure \ref{fig:corner}. The four parameters are constrained at 68\% ($1\sigma$) confidence level to be $\mathrm{log}\phi_0\, \mathrm{[Gpc^{-3}\,yr^{-1}]}\simeq 1.1^{+0.8}_{-1.2}$, $\beta\simeq 0.8^{+2.6}_{-2.9}$, $\gamma\simeq 1.6\pm 0.3$, $\mathrm{log}E_{\rm  max} \,\mathrm{[erg\,Hz^{-1}]} \simeq 34.1^{+1.1}_{-0.7}$. The cumulative rate above certain energy $E \ll E_{\rm max}$ at $z=0$, given by $N(>E) = \phi_0 (E/E_{\rm max})^{1-\gamma}/(\gamma-1)$, is constrained to be $\mathrm{log}N(>10^{32}\rm \, erg\, Hz^{-1})\, \mathrm{[Gpc^{-3}\,yr^{-1}]}\simeq 2.6\pm 0.4$ (68\% confidence). These constraints, although weak due to small-number statistics, broadly agree with our simple analytical results in \S2 and \S3. Our model will provide tighter constraints on the FRB rate function when applied to the much larger samples expected in the near future. 

\begin{figure}
\includegraphics[width=0.48\textwidth,trim=0.0cm 0.0cm 0.0cm 0.0cm]{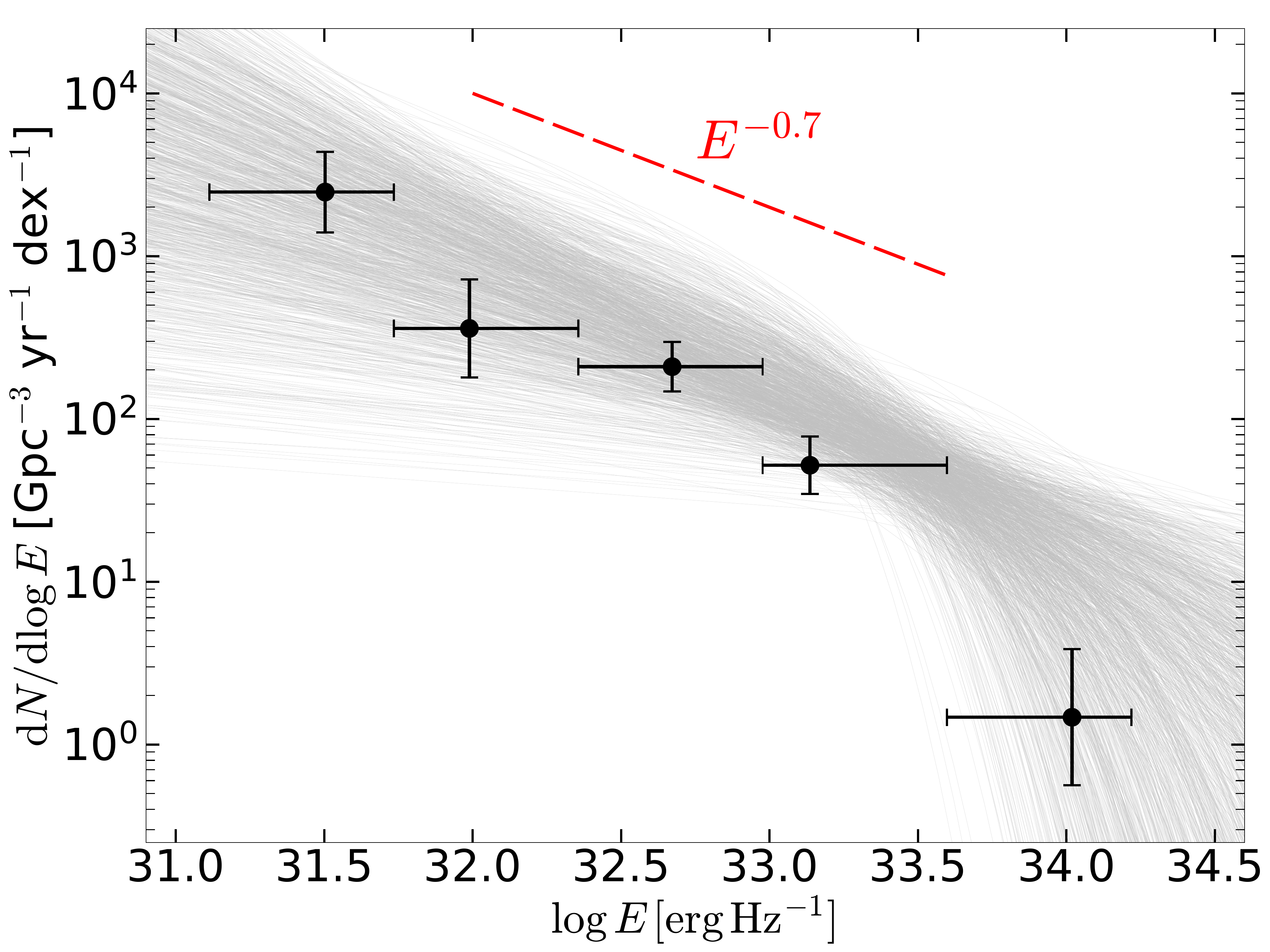}
\caption{The grey curves are the energy distribution functions evaluated at redshift $z=0.5$ (Equation \ref{eq:1}) for $10^3$ randomly selected MCMC samples from our Bayesian analysis. The black circles with (Poissonic) error bars show the results from the standard Schmidt estimator.}
\label{fig:Vmax}
\end{figure}

Finally, we compare the energy distribution function given by the standard ``$1/V_{\rm max}$'' estimator \citep{1968ApJ...151..393S} with our Bayesian results in Figure \ref{fig:Vmax}. We calculate the weighted sum $\sum_i(V_{\mathrm{max}, i}T_{\rm eff})^{-1}$ for each energy bin\footnote{For the small ASKAP sample, the result of the Schmidt estimator is sensitive to binning because of Poisson error. As we group the bursts into logarithmic energy bins, a smooth distribution is obtained when the number of bins is $\leq5$.} ($E$, $E+\Delta E$), where $V_{\mathrm{max}, i}$ is the maximum observable comoving volume for event $i$ in this bin and $T_{\rm eff}$ is the effective all-sky survey time. For the Schmidt estimator, we assume the FRB rate to be independent of redshift, since the redshift evolution is not strongly constrained by the small ASKAP sample. We also show the energy distribution functions (Equation \ref{eq:1}) evaluated at redshift $z=0.5$ (where most bursts are located) for $10^3$ randomly selected MCMC samples of the parameter vector $\b{p}$ from our Bayesian analysis. The reasonable agreement between these two independent methods means that our understanding of the ASKAP FRB statistics is physical.

\section{Sample Completeness}

It is possible that the ASKAP survey has missed some bursts with fluence above our selection threshold $F_{\rm th}=50\rm\, Jy\,ms$, which would cause the sample to be biased. The missing bursts may affect the overall DM and fluence distribution and change the conclusions in the previous sections (which are drawn from a potentially incomplete sample). In this section, we show that the number of missing bursts above $50\rm\, Jy\,ms$ is small ($\lesssim$3) and that the differential incompleteness is a weak function of DM (i.e., the missing bursts do not preferentially have large or small DMs).

% the Bayesian analysis in \S4 needs to include a bias-correction function $f_{\rm bias}(\rm DM, F)$, which accounts for the fraction of bursts missed in each fluence bin ($F$, $F+\Delta F$] as a function of their DMs. Intuitively as seen in Figure \ref{fig:dm}, the differential incompleteness for bursts of different DMs affects the energy distribution slope $\gamma$, and the DM-normalized incompleteness affects the overall volumetric rate normalization $\phi_0$. In this section, we show that the ASKAP sample is reasonably complete above a threshold fluence of $50\rm\, Jy\,ms$, i.e., the bias-correction function $f_{\rm bias}(\rm DM, F)$ is close to unity and is a weak function of DM for any fluence $F>50\rm\, Jy\,ms$.

The signal-to-noise ratio (S/N) for a burst with fluence $F$ and duration $w_{\rm obs}$ can be estimated by \citep{Shannon18}
\begin{equation}
  \label{eq:5}
  \mathrm{S/N} \simeq {F\over w_{\rm obs}} {\sqrt{2B w_{\rm obs}}\over S_{\rm sys}},
\end{equation}
where $B=336\,$MHz is the observing bandwidth, $S_{\rm sys}\simeq 2000\,$Jy is the system equivalent flux density of a single beam, and $w_{\rm obs} = (t_{\rm DM}^2 + t_{\rm samp}^2 + t_{\rm arr}^2)^{1/2}$ is the total duration from a convolution of the DM smearing $t_{\rm DM}$, the sampling time $t_{\rm samp} = 1.26\,$ms, and the burst width at arrival $t_{\rm arr}$ (which is given by the emission width, redshift, and scattering broadening). We have ignored the residual time dispersion $t_{\rm \delta DM}$ due to de-dispersion with a slightly incorrect DM \citep{2003ApJ...596.1142C}. The DM smearing is given by 
\begin{equation}
  \label{eq:6}
  t_{\rm DM} = 3.6 \mathrm{\,ms} {\rm DM_{\rm tot}\over 10^3\, pc\,cm^{-3}} {\Delta \nu\over \rm 1\,MHz} \left(\nu\over 1.32\rm\,GHz\right)^{-3},
\end{equation}
where $\rm DM_{\rm tot}$ is the total DM of a burst, $\Delta \nu =1\,$MHz is the the spectral resolution, and $\nu=1.32\,$GHz is the central frequency of the ASKAP survey. From Equations (\ref{eq:5}) and (\ref{eq:6}), we see that finite temporal and spectral resolution always decreases the S/N by broadening the observed duration $w_{\rm obs}$. 

We caution that Equation (\ref{eq:5}) only provides an approximate estimate of the S/N, because $S_{\rm sys}$ varies with telescope operational status and the burst's localization \citep[see][for detailed discussions]{2019PASA...36....9J}, and some narrow-spectrum bursts only partially fill the bandwidth $B$. Additionally, the burst spectrum may not peak at $\nu=1.32\,$GHz, so the DM smearing may deviate from Equation (\ref{eq:6}) by up to $\sim$$30\%$. Nevertheless, the discussion in this section stays qualitatively true even under these variations.

\begin{figure}
\includegraphics[width=0.48\textwidth,trim=0.0cm 0.0cm 0.0cm 0.0cm]{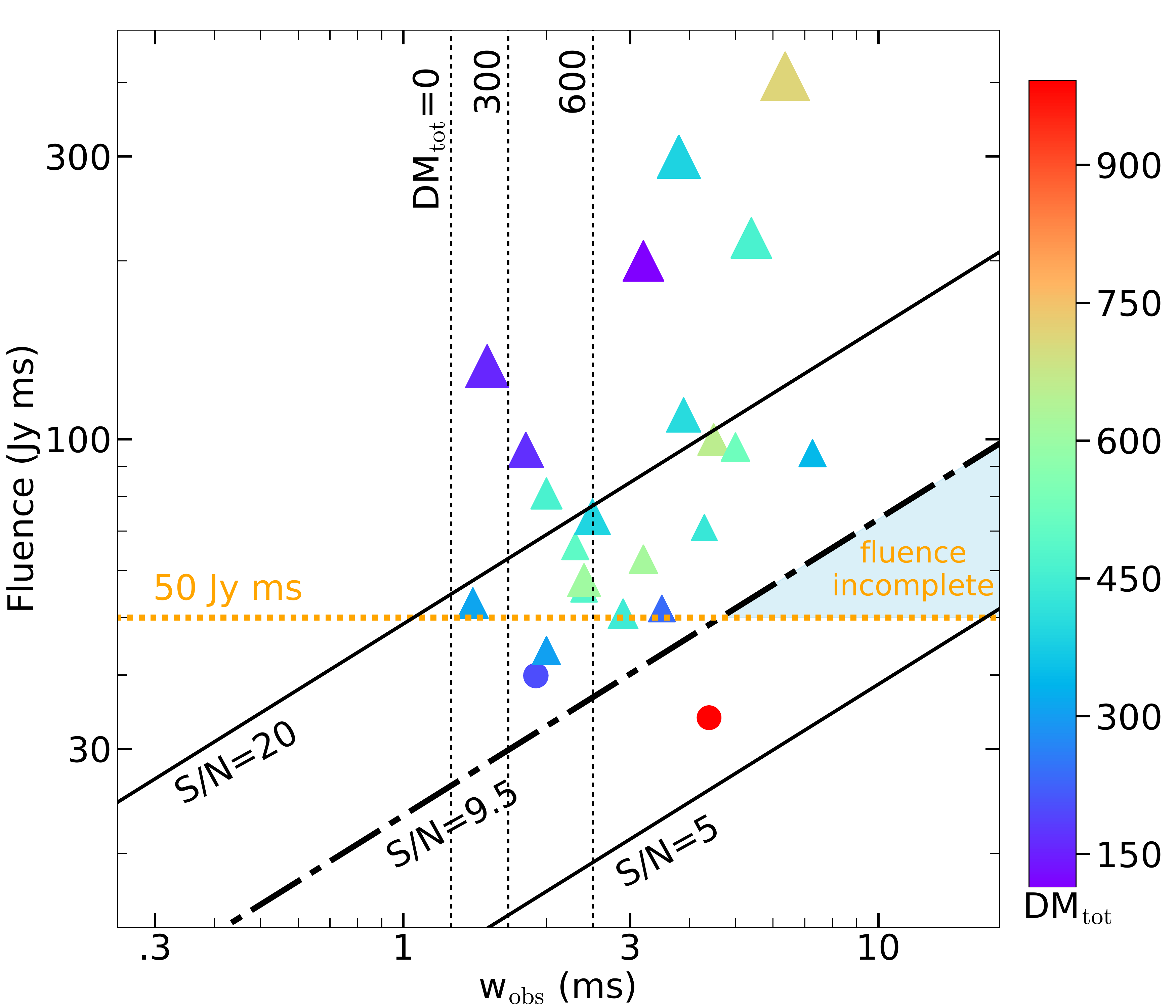}
\caption{The observed durations and fluences for ASKAP FRBs. The colors represent the total DM of the bursts (in $\rm pc\,cm^{-3}$), and symbol sizes are proportional to the S/N of the main beam of detection. Triangles are for FRBs with S/N $>$ 9.5 (the threshold for reliable detection) while circles are for two bursts with S/N $<$ 9.5 (sub-threshold bursts may be recovered by using multi-beam information). We show a few lines of constant S/N = 5 (solid), 9.5 (dash-dotted), and 20 (solid), as given by Equation (\ref{eq:5}). We caution that the S/N for a particular burst may deviate from our analytical estimate due to variations of the system equivalent flux $S_{\rm sys}$ and spectral width $B$. The blue-shaded region marks where the ASKAP sample is fluence-incomplete above $50\rm\,Jy\, ms$. The vertical dotted lines show the minimum observed durations $w_{\rm obs,min}=(t_{\rm samp}^2 + t_{\rm DM}^2)^{1/2}$ (taking $t_{\rm arr}=0$) for three different values of $\rm DM_{\rm tot}=0$, $300$, and $600\rm\, pc\, cm^{-3}$. }
\label{fig:tobs_fluence}
\end{figure}

In Figure \ref{fig:tobs_fluence}, we show the distribution of the observed widths and fluences, following \citet{Keane15}. The observed widths cluster around $\sim$3$\,$ms, because of the sampling time $t_{\rm samp} \sim 1\,$ms and the DM smearing $t_{\rm DM}\sim2\,$ms (such that many shorter bursts with $t_{\rm arr}\lesssim 1\,$ms have been broadened). The longest duration burst has $w_{\rm obs, max}=7.3\,$ms (not strongly affected by DM smearing) and fluence $F = 95\rm\, Jy\, ms$. We assume that the total duration $w_{\rm obs}$ is not correlated with the fluence (no such correlation has been reported), so the number of bursts in any $(w_{\rm obs}, w_{\rm obs}+\Delta w_{\rm obs})$ bin goes as $N(>F)\propto F^{-3/2}$ as shown in Figure \ref{fig:fluence}. Thus, a conservative estimate of the number of missing bursts in the blue-shaded region is given by $2\times(50^{-1.5} - 95^{-1.5})/95^{-1.5}\simeq 3.2$, where we have assumed that all bursts of width near $\sim$7.3 ms and fluence below 95 Jy ms have been missed. Therefore, only a small fraction (about $10\%$) of the bursts above our fluence treshold $F_{\rm th} = 50\rm\, Jy\, ms$ may have been missed in the ASKAP sample. This may be counter-intuitive given the relatively coarse spectral resolution of $\Delta \nu=1\,$MHz. It is the high fluence threshold (only selecting those extremely bright bursts) that pushes the incompleteness region to very long durations where FRBs are sufficiently rare.

% Thus, we should expect roughly equal number of bursts in the two fluence ranges $(50, 80]$ and $(80, \infty)\rm\, Jy\,ms$. Our sample only has three bursts with $w_{\rm obs}>5\,$ms and $F>80\rm\,Jy\,ms$. Additionally, only a fraction of bursts (those longer-duration ones) in the fluence range of $(50, 80]\rm\, Jy\, ms$ are in the blue-shaded region of incompleteness. Therefore, only a small fraction ($\lesssim 10\%$) of the bursts above our fluence treshold $F_{\rm th} = 50\rm\, Jy\, ms$ may have been missed in the ASKAP sample.

\begin{figure}
\includegraphics[width=0.48\textwidth,trim=0.0cm 0.0cm 0.0cm 0.0cm]{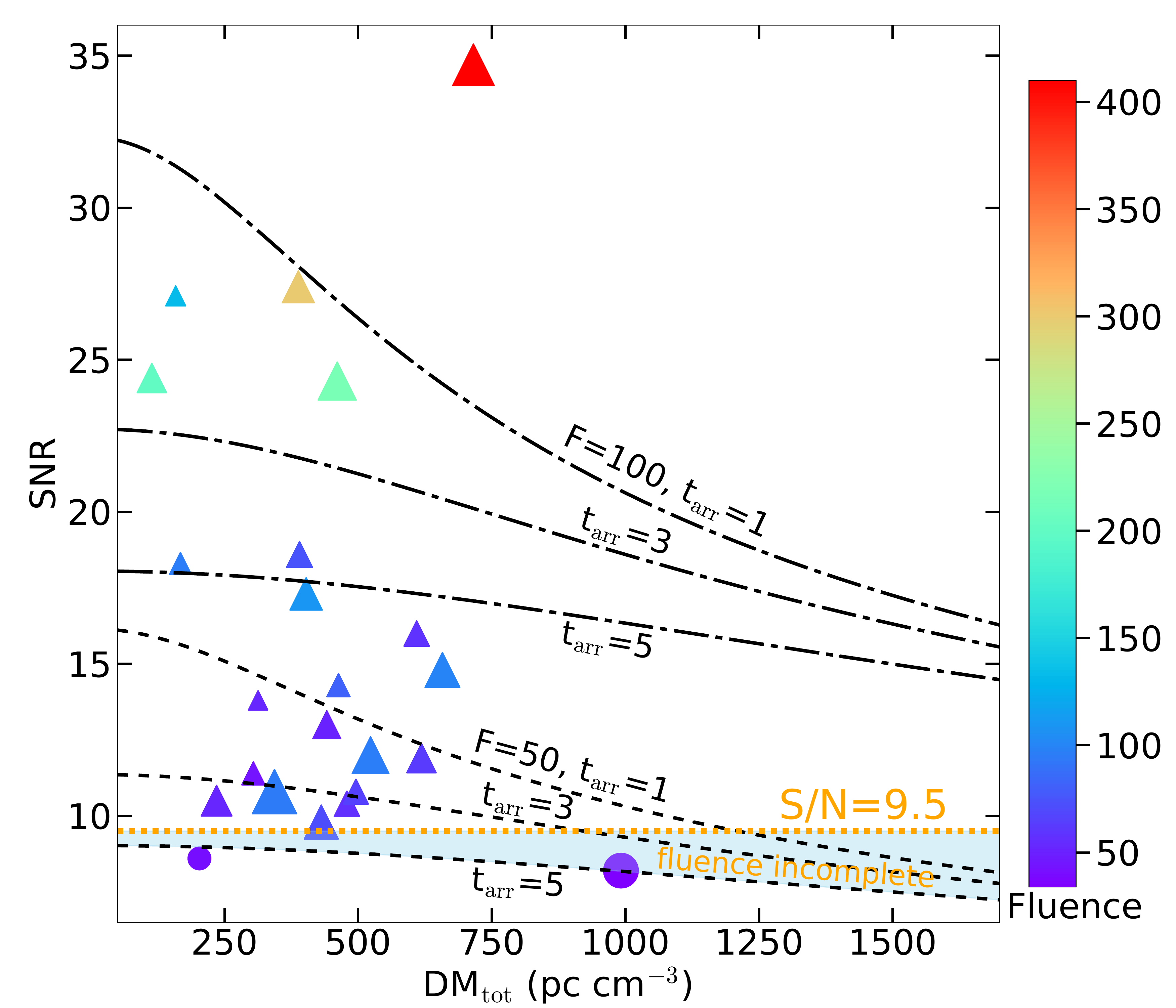}
\caption{The total DM and S/N (of the main beam of detection) for ASKAP FRBs. The colors represent burst fluences, and symbol sizes are proportional to their total duration $w_{\rm obs}$. Triangles are for FRBs with S/N $>$ 9.5 while circles are for two bursts with S/N $<$ 9.5. We also show a few curves of the S/N expected for bursts of a given fluence $F$ (in $\rm\, Jy\, ms$) and arrival width $t_{\rm arr}$ (in $\rm ms$), as given by Equations (\ref{eq:5}) and (\ref{eq:6}). The dashed-dotted curves are for $F=100$ and the dashed curves are for $F=50$. The variation $\rm{S/N}(DM)$ in each curve is only due to DM smearing $t_{\rm DM}$. The $(F,\ t_{\rm arr})=\rm (50, 1)$ curve represents short-duration faint bursts which should be well detected for the DM range of our sample. The $(F,\ t_{\rm arr})=\rm (50, 5)$ curve represents long-duration faint bursts which are likely missed in our sample. The blue-shaded region marks where our sample is fluence-incomplete above $50\rm\,Jy\, ms$. The shallow slope of the $(F,\ t_{\rm arr})=\rm (50, 5)$ means that the differential incompleteness is a weak function of DM for these long-duration bursts.}
\label{fig:dm_snr}
\end{figure}

Figure \ref{fig:dm_snr} shows the total DM and S/N of the main beam of detection for each burst. We also show a few curves of the S/N expected for bursts of a given fluence $F$ (in $\rm\, Jy\, ms$) and arrival width $t_{\rm arr}$ (in $\rm ms$). The variation $\rm{S/N}(DM)$ in each curve is only due to DM smearing $t_{\rm DM}$. We find that (i) long-duration faint bursts, represented by the $(F,\ t_{\rm arr}) = (50, 5)$ curve, are likely missed in our sample; and (ii) for long-duration bursts, represented by the two curves with $t_{\rm arr}=5$, the S/N depends weakly on $\rm DM_{\rm tot}$, because the total duration $w_{\rm obs}$ is dominated by the arrival width $t_{\rm arr}$ ($\gg t_{\rm DM}$). Thus, even if some long-duration faint bursts have been missed in our sample, 
the differential incompleteness\footnote{In the future, when a better understanding of the arrival durations ($t_{\rm arr}$, including scattering broadening) is available, it is straightforward to include a bias correction factor $f_{\rm bias}(\rm DM)$ for bursts of different DMs in our Bayesian analysis.} is a weak function of DM (i.e., the missing bursts do not preferentially have large or small DMs). The abrupt cutoff in the number of bursts with $\rm DM_{\rm tot}\gtrsim 750\rm\, pc\,cm^{-3}$ at $\rm S/N>9.5$ and $F> 50\rm\, Jy\, ms$ is not caused by loss of S/N due to excessive DM smearing. Instead, it can only be due to the lack of these events arriving at the telescope, which is explained by a cutoff in the volumetric rate above a maximum burst energy $E_{\rm max}$ in our model.

% The following conclusions can be drawn. (i) Short-duration faint bursts, represented by the $(F,\ t_{\rm arr})=\rm (50, 1)$ curve, are well detected for the DM range of our sample (only those with $\rm DM_{\rm tot}\gtrsim 1200$ may be sufficiently broadened by DM smearing to cause non-detection); (ii) Bright bursts, represented by the three dash-dotted curves with $F=100$, are always well detected regardless of their DMs;

Recently, \citet{2019arXiv190500755C} studied how the finite time and frequency resolution affects the distributions of the observed burst duration, DM, and flux. The author argued that a large fraction of FRBs with short durations $t_{\rm arr}\ll 1\,$ms and high DMs may have been missed in current surveys, including ASKAP. Our approach is different in that we aim to have a fluence-complete sample above a threshold S/N whereas \citet{2019arXiv190500755C} focused on the observed bursts above a threshold detection flux (see their Equation 16). We know that $\mathrm{S/N}\propto Fw_{\rm obs}^{-1/2}\propto S w_{\rm obs}^{1/2}$ (Equation \ref{eq:5}), where $S = F/w_{\rm obs}=S_0t_{\rm arr}/w_{\rm obs}$ is the detection flux and $S_0$ is the ``intrinsic'' flux arriving at the telescope. Thus, fluence completeness depends crucially on the number of longest-duration bursts whereas flux completeness relies on shortest-duration ones. If there exists a large number of short bursts with $t_{\rm arr}\ll 1\,$ms (possibly down to nano-second timescale), as suggested by the Parkes sample \citep{Ravi19} and giant pulses of Galactic pulsars, it is extremely difficult to achieve a flux($S_0$)-complete sample. This is the motivation for why we choose to study the energy distribution function of FRBs but not the luminosity function.

\section{Discussion}
We note a few caveats and highlight some implications of our study.

% (1) First, incompleteness becomes significant for pulses narrower than the time resolution. Moreover, the FRB-search algorithm loses sensitivity when the dispersion delay across a spectral channel width is much greater (for high $\mathrm{DM}$) or much smaller (low $\mathrm{DM}$) than the time resolution. Long duration ($\gtrsim 10\,$ms) faint bursts are likely missed due to low S/N. These biases are all related to the distribution of pulse durations (and possible scattering broadening), which we currently do not have a model for. In the future, our Bayesian analysis can be augmented by a correction function $f_{\rm in} (\mathrm{DM})$, which accounts for the differential incompleteness for bursts of different $\mathrm{DM}$'s.

(1) Instead of taking $F_{\rm th}=50\rm\,Jy\,ms$, we have also tested a more conservative choice of fluence threshold $F_{\rm th}=60\rm\,Jy\,ms$ (which includes 15 out of the 23 bursts). According to \S5, this smaller sample is better fluence-complete with less bias against long-duration faint bursts. We find that our model still provides a good fit to the DM and fluence distributions, although the statistical constraints on the model parameters are slightly worse.

(2) We have assumed that the DM excess (beyond the Milky Way's contribution) is largely due to the intergalactic medium and hence can be used as a distance indicator. However, as we know from FRB 121102 \citep{Tendulkar17}, the host galaxy and the circum-burst medium may contribute a significant fraction of DM. Unfortunately, the properties of FRB host galaxies are still highly uncertain. We also tried subtracting a constant host-galaxy contribution of $30\rm\,pc\,cm^{-3}$. The resulting distribution $N(<{\rm DM})$ becomes slightly shallower on the low-DM end, and the Bayesian analysis gives similar parameter constraints. In the future, with a larger FRB sample, it is possible to constrain the (averaged) host-galaxy and circum-burst DM \citep[e.g., from a supernova remnant,][]{Connor16,Piro16,Piro17,Piro18} by studying the deviation of the DM distribution from a power-law on the low-DM end. On the other hand, FRBs at $z\gtrsim 0.8$ may randomly have their sight lines intersecting with a few galactic haloes \citep{McQuinn14, Prochaska19}. This effect causes stochastic deviations of the DM distribution from our model.
% Taking this effect into account will make the distribution $N(<{\rm DM})$ steeper and it goes in the direction of giving a slightly smaller energy distribution index $\gamma$.

% (4) In Figure \ref{fig:dm_fluence}, there is an outlier (FRB 180110) with inferred energy higher than $E_{\rm max}$ in Equation (\ref{eq:Emax}) by a factor of 5. This can be explained by a combination of a slightly larger $E_{\rm max}$ and that an exponential cutoff allows burst energies up to a few times above $E_{\rm max}$.
% % A more realistic energy distribution function may have an exponential drop above the maximum energy (instead of an abrupt cutoff in our simple model).
% % This may be expected model in Equation (\ref{eq:1}), a small fraction bursts may have energies up to a few times higher than $E_{\rm max}$.
% Alternatively, this outlier may have a large DM contribution from a combination of its host galaxy, local circum-burst environment, and intervening haloes. Magnification by plasma lensing is also a possibility \citep{Cordes17}, although the fact that most FRBs adhere to the limit we derive argues that such effects are small in most cases.

(3) The scaling in Equation (\ref{eq:fluence}) breaks down at sufficiently small fluences.
% because most bursts would come from very high redshifts where the spacetime is no-longer Euclidean (the co-moving volume increases much slower than $D^3$) and there could be strong redshift evolution of the rate normalization.
This is because (i) the spacetime is no-longer Euclidean at high redshift (the co-moving volume increases much slower than $z^3$), and (ii) $N(>F)$ is no longer dominated by bursts with $E\sim E_{\rm max}$ when the majority of them in the Universe have already been counted at higher fluences. For instance, at $F\lesssim E_{\rm max}/4\pi D^2_{\rm L}(z=5)\simeq 1\rm\,Jy\,ms$, the cumulative number $N(>F)$ should be dominated by bursts with $E\ll E_{\rm max}$, and hence the fluence distribution becomes $N(>F)\propto F^{1-\gamma}$. This means that a sensitive telescope with fluence threshold $\ll 1\rm\,Jy\,ms$ can directly measure the energy distribution slope $\gamma$ by source counting. This may explain the difference in logN-logF slopes given by the Parkes and ASKAP samples \citep{James19}, since the former has a much lower fluence threshold. In the future, with a much larger (well-controlled) sample down to a lower fluence threshold, it is possible to constrain the redshift evolution of FRB rate \citep{Macquart18}, which will in turn constrain the progenitor models for FRBs.

(4) The first repeater has been localized to a metal poor star-forming dwarf galaxy, suggesting that there may be some relation to a long gamma-ray bursts and superluminous supernovae \citep{Metzger17}. As more FRBs are localized, an important constraint to the progenitor model will be how much variety is seen in the hosts.

If, for example, some FRBs are found to have distinct hosts from others, one might conclude that they have different progenitors. If the distributions of the various FRB populations continue to obey similar energy distributions and maximum energies as we describe here, it argues that the underlying source is the same even if the situation that generated the source is different. For instance, in the magnetar picture, one could imagine magnetars generated both from young massive stars and old stellar environments by neutron star or white dwarf mergers. FRBs from these two populations could vary in a number of ways, from the host galaxies, to the local DM, to the rotation measure and presence of a persistent radio source. But by looking at the burst statistics as we describe here, we may understand if the underlying source is the same.

\section{Summary}

This work uses the ASKAP sample, which we show to be reasonably complete above $50\rm\, Jy\,ms$, to study the energetics and cosmological rate of the whole FRB population.

We find that the ASKAP sample is well described with a power-law energy distribution with slope $\gamma = -d{\rm log}N/d {\rm log}E \sim 1.7$, independent of whether they individually repeat themselves. This is because the observed DM distribution scales as \mbox{$N(>{\rm DM})\propto \mathrm{DM}^{5-2\gamma}$} (for $\gamma>1$). The abrupt cutoff in the number of bursts with $\rm DM\gtrsim 750\rm\, pc\,cm^{-3}$ suggests that the FRB population has a maximum specific energy of $E_{\rm max}\sim \mathrm{a\, few}\times10^{33}\rm\, erg\, Hz^{-1}$ above which the  volumetric rate rapidly drops. For a spectral width of $\sim$$1\,$GHz, this implies a maximum isotropic energy of order $10^{42}\rm\,erg$ or isotropic luminosity of order $10^{45}\rm\,erg \,s^{-1}$ for millisecond duration, which is a factor of a few to ten higher than that found by \citet{Luo18} from the Parkes sample. 

The existence of a maximum energy $E_{\rm max}$ causes the observed fluence distribution to be $N(>F)\propto F^{-3/2}$ at high fluences, because the number of bursts is dominated by those at distances $D \sim \sqrt{E_{\rm max}/4\pi F}$. At sufficiently low fluences, the distribution approaches $N(>F)\propto F^{1-\gamma}$, reflecting the intrinsic energy distribution. The transition between these two regimes is sensitive to the evolution of FRB rate at high redshift.

Besides simple analytical insights, we also present a numerical Bayesian analysis of the ASKAP sample by comparing the observed distributions with the predictions from a Schechter-like model for the FRB rate (Equation \ref{eq:1}). The results are shown in Figure \ref{fig:corner}. The energy distribution power-law index and the maximum energy are constrained to be $\gamma\simeq 1.6\pm 0.3$ and $\mathrm{log}E_{\rm  max} \,\mathrm{[erg\,Hz^{-1}]} \simeq 34.1^{+1.1}_{-0.7}$ (68\% confidence), respectively. From the survey exposure time, we further infer a cumulative local volumetric rate of $\mathrm{log}N(E>10^{32}\rm \, erg\, Hz^{-1})\, \mathrm{[Gpc^{-3}\,yr^{-1}]}\simeq 2.6\pm 0.4$ (68\% confidence). Finally, we compare the energy distribution function given by the standard ``$1/V_{\rm max}$'' estimator with that from our Bayesian approach. These two independent methods give consistent results as shown in Figure \ref{fig:Vmax}, which means that our understanding of the ASKAP FRB statistics is physical. Our model will give tighter constraints on the statistical properties of FRB rate when applied to the much larger samples to be collected in the near future.

% Our fitting functions (Equations \ref{eq:fit} and \ref{eq:NgtF}) can be directly applied to the brightest portion of other FRB samples in the near future. It is also straightforward to generalize our model to include cosmological effects and then to study how the burst distribution, energetics, and rates evolve with redshift.

\acknowledgments
We thank the two anonymous referees for carefully reading the manuscript and providing valuable criticisms and suggestions. We thank Jonathan Katz, Liam Connor, and Ue-Li Pen for useful comments. WL is supported by the David and Ellen Lee Fellowship at Caltech.

\bibliographystyle{apj}
\bibliography{refs}

\begin{thebibliography}{73}
\expandafter\ifx\csname natexlab\endcsname\relax\def\natexlab#1{#1}\fi

\bibitem[{{Aschwanden} {et~al.}(2016){Aschwanden}, {Crosby}, {Dimitropoulou},
  {Georgoulis}, {Hergarten}, {McAteer}, {Milovanov}, {Mineshige}, {Morales},
  {Nishizuka}, {Pruessner}, {Sanchez}, {Sharma}, {Strugarek}, \&
  {Uritsky}}]{Aschwanden16}
{Aschwanden}, M.~J., {Crosby}, N.~B., {Dimitropoulou}, M., {et~al.} 2016, \ssr,
  198, 47

\bibitem[{{Bailes} {et~al.}(2017){Bailes}, {Jameson}, {Flynn}, {Bateman},
  {Barr}, {Bhandari}, {Bunton}, {Caleb}, {Campbell-Wilson}, {Farah},
  {Gaensler}, {Green}, {Hunstead}, {Jankowski}, {Keane}, {Krishnan}, {Murphy},
  {O'Neill}, {Os{\l}owski}, {Parthasarathy}, {Ravi}, {Rosado}, \&
  {Temby}}]{Bailes17}
{Bailes}, M., {Jameson}, A., {Flynn}, C., {et~al.} 2017, Publications of the
  Astronomical Society of Australia, 34, e045

\bibitem[{{Bak} {et~al.}(1987){Bak}, {Tang}, \& {Wiesenfeld}}]{Bak87}
{Bak}, P., {Tang}, C., \& {Wiesenfeld}, K. 1987, \prl, 59, 381

\bibitem[{{Bannister} {et~al.}(2017){Bannister}, {Shannon}, {Macquart},
  {Flynn}, {Edwards}, {O'Neill}, {Os{\l}owski}, {Bailes}, {Zackay}, {Clarke},
  {D'Addario}, {Dodson}, {Hall}, {Jameson}, {Jones}, {Navarro}, {Trinh},
  {Allison}, {Anderson}, {Bell}, {Chippendale}, {Collier}, {Heald}, {Heywood},
  {Hotan}, {Lee-Waddell}, {Madrid}, {Marvil}, {McConnell}, {Popping},
  {Voronkov}, {Whiting}, {Allen}, {Bock}, {Brodrick}, {Cooray}, {DeBoer},
  {Diamond}, {Ekers}, {Gough}, {Hampson}, {Harvey-Smith}, {Hay}, {Hayman},
  {Jackson}, {Johnston}, {Koribalski}, {McClure-Griffiths}, {Mirtschin}, {Ng},
  {Norris}, {Pearce}, {Phillips}, {Roxby}, {Troup}, \&
  {Westmeier}}]{Bannister17}
{Bannister}, K.~W., {Shannon}, R.~M., {Macquart}, J.~P., {et~al.} 2017, \apj,
  841, L12

\bibitem[{{Beloborodov}(2017)}]{Beloborodov17}
{Beloborodov}, A.~M. 2017, \apjl, 843, L26

\bibitem[{{Bhandari} {et~al.}(2018){Bhandari}, {Keane}, {Barr}, {Jameson},
  {Petroff}, {Johnston}, {Bailes}, {Bhat}, {Burgay}, {Burke-Spolaor}, {Caleb},
  {Eatough}, {Flynn}, {Green}, {Jankowski}, {Kramer}, {Krishnan}, {Morello},
  {Possenti}, {Stappers}, {Tiburzi}, {van Straten}, {Andreoni}, {Butterley},
  {Chand ra}, {Cooke}, {Corongiu}, {Coward}, {Dhillon}, {Dodson}, {Hardy},
  {Howell}, {Jaroenjittichai}, {Klotz}, {Littlefair}, {Marsh}, {Mickaliger},
  {Muxlow}, {Perrodin}, {Pritchard}, {Sawangwit}, {Terai}, {Tominaga}, {Torne},
  {Totani}, {Trois}, {Turpin}, {Niino}, {Wilson}, {Albert}, {Andr{\'e}},
  {Anghinolfi}, {Anton}, {Ardid}, {Aubert}, {Avgitas}, {Baret},
  {Barrios-Mart{\'\i}}, {Basa}, {Belhorma}, {Bertin}, {Biagi}, {Bormuth},
  {Bourret}, {Bouwhuis}, {Br{\^a}nza{\c{s}}}, {Bruijn}, {Brunner}, {Busto},
  {Capone}, {Caramete}, {Carr}, {Celli}, {Moursli}, {Chiarusi}, {Circella},
  {Coelho}, {Coleiro}, {Coniglione}, {Costantini}, {Coyle}, {Creusot},
  {D{\'\i}az}, {Deschamps}, {De Bonis}, {Distefano}, {Palma}, {Domi},
  {Donzaud}, {Dornic}, {Drouhin}, {Eberl}, {Bojaddaini}, {Khayati},
  {Els{\"a}sser}, {Enzenh{\"o}fer}, {Ettahiri}, {Fassi}, {Felis}, {Fusco},
  {Gay}, {Giordano}, {Glotin}, {Gregoire}, {Gracia-Ruiz}, {Graf}, {Hallmann},
  {van Haren}, {Heijboer}, {Hello}, {Hern{\'a}ndez-Rey}, {H{\"o}{\ss}l},
  {Hofest{\"a}dt}, {Hugon}, {Illuminati}, {James}, {de Jong}, {Jongen},
  {Kadler}, {Kalekin}, {Katz}, {Kie{\ss}ling}, {Kouchner}, {Kreter},
  {Kreykenbohm}, {Kulikovskiy}, {Lachaud}, {Lahmann}, {Lef{\`e}vre}, {Leonora},
  {Loucatos}, {Marcelin}, {Margiotta}, {Marinelli}, {Mart{\'\i}nez-Mora},
  {Mele}, {Melis}, {Michael}, {Migliozzi}, {Moussa}, {Navas}, {Nezri},
  {Organokov}, {P{\v{a}}v{\v{a}}la{\c{s}}}, {Pellegrino}, {Perrina},
  {Piattelli}, {Popa}, {Pradier}, {Quinn}, {Racca}, {Riccobene},
  {S{\'a}nchez-Losa}, {Salda{\~n}a}, {Salvadori}, {Samtleben}, {Sanguineti},
  {Sapienza}, {Sch{\"u}ssler}, {Sieger}, {Spurio}, {Stolarczyk}, {Taiuti},
  {Tayalati}, {Trovato}, {Turpin}, {T{\"o}nnis}, {Vallage}, {Van Elewyck},
  {Versari}, {Vivolo}, {Vizzocca}, {Wilms}, {Zornoza}, \&
  {Z{\'u}{\~n}iga}}]{Bhandari18}
{Bhandari}, S., {Keane}, E.~F., {Barr}, E.~D., {et~al.} 2018, \mnras, 475, 1427

\bibitem[{{Bhattacharya} {et~al.}(2019){Bhattacharya}, {Kumar}, \&
  {Lorimer}}]{2019arXiv190210225B}
{Bhattacharya}, M., {Kumar}, P., \& {Lorimer}, D. 2019, arXiv e-prints,
  arXiv:1902.10225

\bibitem[{{Burke-Spolaor} \& {Bannister}(2014)}]{Burke-Spolaor14}
{Burke-Spolaor}, S., \& {Bannister}, K.~W. 2014, \apj, 792, 19

\bibitem[{{Caleb} {et~al.}(2017){Caleb}, {Flynn}, {Bailes}, {Barr}, {Bateman},
  {Bhandari}, {Campbell-Wilson}, {Farah}, {Green}, {Hunstead}, {Jameson},
  {Jankowski}, {Keane}, {Parthasarathy}, {Ravi}, {Rosado}, {van Straten}, \&
  {Venkatraman Krishnan}}]{Caleb17}
{Caleb}, M., {Flynn}, C., {Bailes}, M., {et~al.} 2017, \mnras, 468, 3746

\bibitem[{{Champion} {et~al.}(2016){Champion}, {Petroff}, {Kramer}, {Keith},
  {Bailes}, {Barr}, {Bates}, {Bhat}, {Burgay}, {Burke-Spolaor}, {Flynn},
  {Jameson}, {Johnston}, {Ng}, {Levin}, {Possenti}, {Stappers}, {van Straten},
  {Thornton}, {Tiburzi}, \& {Lyne}}]{Champion16}
{Champion}, D.~J., {Petroff}, E., {Kramer}, M., {et~al.} 2016, \mnras, 460, L30

\bibitem[{{Chatterjee} {et~al.}(2017){Chatterjee}, {Law}, {Wharton},
  {Burke-Spolaor}, {Hessels}, {Bower}, {Cordes}, {Tendulkar}, {Bassa},
  {Demorest}, {Butler}, {Seymour}, {Scholz}, {Abruzzo}, {Bogdanov}, {Kaspi},
  {Keimpema}, {Lazio}, {Marcote}, {McLaughlin}, {Paragi}, {Ransom}, {Rupen},
  {Spitler}, \& {van Langevelde}}]{Chatterjee17}
{Chatterjee}, S., {Law}, C.~J., {Wharton}, R.~S., {et~al.} 2017, \nat, 541, 58

\bibitem[{{CHIME/FRB Collaboration} {et~al.}(2018){CHIME/FRB Collaboration},
  {Amiri}, {Bandura}, {Berger}, {Bhardwaj}, {Boyce}, {Boyle}, {Brar},
  {Burhanpurkar}, {Chawla}, {Chowdhury}, {Cliche}, {Cranmer}, {Cubranic},
  {Deng}, {Denman}, {Dobbs}, {Fandino}, {Fonseca}, {Gaensler}, {Giri},
  {Gilbert}, {Good}, {Guliani}, {Halpern}, {Hinshaw}, {H{\"o}fer}, {Josephy},
  {Kaspi}, {Landecker}, {Lang}, {Liao}, {Masui}, {Mena-Parra}, {Naidu},
  {Newburgh}, {Ng}, {Patel}, {Pen}, {Pinsonneault-Marotte}, {Pleunis}, {Rafiei
  Ravandi}, {Ransom}, {Renard}, {Scholz}, {Sigurdson}, {Siegel}, {Smith},
  {Stairs}, {Tendulkar}, {Vand erlinde}, \& {Wiebe}}]{Amiri18}
{CHIME/FRB Collaboration}, {Amiri}, M., {Bandura}, K., {et~al.} 2018, \apj,
  863, 48

\bibitem[{{Connor}(2019)}]{2019arXiv190500755C}
{Connor}, L. 2019, arXiv e-prints, arXiv:1905.00755

\bibitem[{{Connor} {et~al.}(2016){Connor}, {Sievers}, \& {Pen}}]{Connor16}
{Connor}, L., {Sievers}, J., \& {Pen}, U.-L. 2016, \mnras, 458, L19

\bibitem[{{Cordes} \& {Lazio}(2002)}]{2002astro.ph..7156C}
{Cordes}, J.~M., \& {Lazio}, T.~J.~W. 2002, arXiv e-prints, astro

\bibitem[{{Cordes} \& {McLaughlin}(2003)}]{2003ApJ...596.1142C}
{Cordes}, J.~M., \& {McLaughlin}, M.~A. 2003, \apj, 596, 1142

\bibitem[{{Fasano} \& {Franceschini}(1987)}]{1987MNRAS.225..155F}
{Fasano}, G., \& {Franceschini}, A. 1987, \mnras, 225, 155

\bibitem[{Foreman-Mackey(2016)}]{corner}
Foreman-Mackey, D. 2016, The Journal of Open Source Software, 24

\bibitem[{{Gourdji} {et~al.}(2019){Gourdji}, {Michilli}, {Spitler}, {Hessels},
  {Seymour}, {Cordes}, \& {Chatterjee}}]{2019ApJ...877L..19G}
{Gourdji}, K., {Michilli}, D., {Spitler}, L.~G., {et~al.} 2019, \apj, 877, L19

\bibitem[{{James} {et~al.}(2019{\natexlab{a}}){James}, {Ekers}, {Macquart},
  {Bannister}, \& {Shannon}}]{James19}
{James}, C.~W., {Ekers}, R.~D., {Macquart}, J.~P., {Bannister}, K.~W., \&
  {Shannon}, R.~M. 2019{\natexlab{a}}, \mnras, 483, 1342

\bibitem[{{James} {et~al.}(2019{\natexlab{b}}){James}, {Bannister}, {Macquart},
  {Ekers}, {Oslowski}, {Shannon}, {Allison}, {Chippendale}, {Collier},
  {Franzen}, {Hotan}, {Leach}, {McConnell}, {Pilawa}, {Voronkov}, \&
  {Whiting}}]{2019PASA...36....9J}
{James}, C.~W., {Bannister}, K.~W., {Macquart}, J.~P., {et~al.}
  2019{\natexlab{b}}, \pasa, 36, e009

\bibitem[{{Jauncey}(1967)}]{1967Natur.216..877J}
{Jauncey}, D.~L. 1967, \nat, 216, 877

\bibitem[{{Katz}(1986)}]{Katz86}
{Katz}, J.~I. 1986, \jgr, 91, 10

\bibitem[{{Katz}(2016)}]{Katz16}
---. 2016, \apj, 826, 226

\bibitem[{{Keane} \& {Petroff}(2015)}]{Keane15}
{Keane}, E.~F., \& {Petroff}, E. 2015, \mnras, 447, 2852

\bibitem[{{Kulkarni} {et~al.}(2014){Kulkarni}, {Ofek}, {Neill}, {Zheng}, \&
  {Juric}}]{Kulkarni14}
{Kulkarni}, S.~R., {Ofek}, E.~O., {Neill}, J.~D., {Zheng}, Z., \& {Juric}, M.
  2014, \apj, 797, 70

\bibitem[{{Kumar} {et~al.}(2017){Kumar}, {Lu}, \& {Bhattacharya}}]{Kumar17}
{Kumar}, P., {Lu}, W., \& {Bhattacharya}, M. 2017, \mnras, 468, 2726

\bibitem[{{Law} {et~al.}(2017){Law}, {Abruzzo}, {Bassa}, {Bower},
  {Burke-Spolaor}, {Butler}, {Cantwell}, {Carey}, {Chatterjee}, {Cordes},
  {Demorest}, {Dowell}, {Fender}, {Gourdji}, {Grainge}, {Hessels}, {Hickish},
  {Kaspi}, {Lazio}, {McLaughlin}, {Michilli}, {Mooley}, {Perrott}, {Ransom},
  {Razavi-Ghods}, {Rupen}, {Scaife}, {Scott}, {Scholz}, {Seymour}, {Spitler},
  {Stovall}, {Tendulkar}, {Titterington}, {Wharton}, \& {Williams}}]{Law17}
{Law}, C.~J., {Abruzzo}, M.~W., {Bassa}, C.~G., {et~al.} 2017, \apj, 850, 76

\bibitem[{{Li} {et~al.}(2013){Li}, {Nan}, \& {Pan}}]{Li13}
{Li}, D., {Nan}, R., \& {Pan}, Z. 2013, in IAU Symposium, Vol. 291, Neutron
  Stars and Pulsars: Challenges and Opportunities after 80 years, ed. J.~{van
  Leeuwen}, 325--330

\bibitem[{{Li} {et~al.}(2019){Li}, {Yalinewich}, \& {Breysse}}]{Li19}
{Li}, D., {Yalinewich}, A., \& {Breysse}, P.~C. 2019, arXiv e-prints

\bibitem[{{Li} {et~al.}(2017){Li}, {Huang}, {Zhang}, {Li}, \&
  {Li}}]{2017RAA....17....6L}
{Li}, L.-B., {Huang}, Y.-F., {Zhang}, Z.-B., {Li}, D., \& {Li}, B. 2017,
  Research in Astronomy and Astrophysics, 17, 6

\bibitem[{{Lorimer} {et~al.}(2007){Lorimer}, {Bailes}, {McLaughlin},
  {Narkevic}, \& {Crawford}}]{Lorimer07}
{Lorimer}, D.~R., {Bailes}, M., {McLaughlin}, M.~A., {Narkevic}, D.~J., \&
  {Crawford}, F. 2007, Science, 318, 777

\bibitem[{{Lu} \& {Kumar}(2016)}]{Lu16}
{Lu}, W., \& {Kumar}, P. 2016, \mnras, 461, L122

\bibitem[{{Lu} \& {Kumar}(2019)}]{lu18}
---. 2019, \mnras, 483, L93

\bibitem[{{Luo} {et~al.}(2018){Luo}, {Lee}, {Lorimer}, \& {Zhang}}]{Luo18}
{Luo}, R., {Lee}, K., {Lorimer}, D.~R., \& {Zhang}, B. 2018, \mnras, 481, 2320

\bibitem[{{Lyubarsky}(2014)}]{Lyubarsky14}
{Lyubarsky}, Y. 2014, \mnras, 442, L9

\bibitem[{{Maan} \& {van Leeuwen}(2017)}]{Maan17}
{Maan}, Y., \& {van Leeuwen}, J. 2017, arXiv e-prints, arXiv:1709.06104

\bibitem[{{Macquart} \& {Ekers}(2018)}]{Macquart18}
{Macquart}, J.~P., \& {Ekers}, R. 2018, \mnras, 480, 4211

\bibitem[{{Macquart} {et~al.}(2019){Macquart}, {Shannon}, {Bannister}, {James},
  {Ekers}, \& {Bunton}}]{2019ApJ...872L..19M}
{Macquart}, J.~P., {Shannon}, R.~M., {Bannister}, K.~W., {et~al.} 2019, \apj,
  872, L19

\bibitem[{{McQuinn}(2014)}]{McQuinn14}
{McQuinn}, M. 2014, \apj, 780, L33

\bibitem[{{Metzger} {et~al.}(2017){Metzger}, {Berger}, \&
  {Margalit}}]{Metzger17}
{Metzger}, B.~D., {Berger}, E., \& {Margalit}, B. 2017, \apj, 841, 14

\bibitem[{{Nicholl} {et~al.}(2017){Nicholl}, {Williams}, {Berger}, {Villar},
  {Alexander}, {Eftekhari}, \& {Metzger}}]{Nicholl17}
{Nicholl}, M., {Williams}, P.~K.~G., {Berger}, E., {et~al.} 2017, \apj, 843, 84

\bibitem[{{Oppermann} {et~al.}(2018){Oppermann}, {Yu}, \&
  {Pen}}]{2018MNRAS.475.5109O}
{Oppermann}, N., {Yu}, H.-R., \& {Pen}, U.-L. 2018, \mnras, 475, 5109

\bibitem[{{Patel} {et~al.}(2018){Patel}, {Agarwal}, {Bhardwaj}, {Boyce},
  {Brazier}, {Chatterjee}, {Chawla}, {Kaspi}, {Lorimer}, {McLaughlin},
  {Parent}, {Pleunis}, {Ransom}, {Scholz}, {Wharton}, {Zhu}, {Alam}, {Caballero
  Valdez}, {Camilo}, {Cordes}, {Crawford}, {Deneva}, {Ferdman}, {Freire},
  {Hessels}, {Nguyen}, {Stairs}, {Stovall}, \& {van Leeuwen}}]{Patel18}
{Patel}, C., {Agarwal}, D., {Bhardwaj}, M., {et~al.} 2018, \apj, 869, 181

\bibitem[{{Peacock}(1983)}]{1983MNRAS.202..615P}
{Peacock}, J.~A. 1983, \mnras, 202, 615

\bibitem[{{Pen} \& {Connor}(2015)}]{Pen15}
{Pen}, U.-L., \& {Connor}, L. 2015, \apj, 807, 179

\bibitem[{{Petroff} {et~al.}(2019{\natexlab{a}}){Petroff}, {Hessels}, \&
  {Lorimer}}]{2019A&ARv..27....4P}
{Petroff}, E., {Hessels}, J.~W.~T., \& {Lorimer}, D.~R. 2019{\natexlab{a}},
  \aapr, 27, 4

\bibitem[{{Petroff} {et~al.}(2014){Petroff}, {van Straten}, {Johnston},
  {Bailes}, {Barr}, {Bates}, {Bhat}, {Burgay}, {Burke-Spolaor}, {Champion},
  {Coster}, {Flynn}, {Keane}, {Keith}, {Kramer}, {Levin}, {Ng}, {Possenti},
  {Stappers}, {Tiburzi}, \& {Thornton}}]{Petroff14}
{Petroff}, E., {van Straten}, W., {Johnston}, S., {et~al.} 2014, \apj, 789, L26

\bibitem[{{Petroff} {et~al.}(2016){Petroff}, {Barr}, {Jameson}, {Keane},
  {Bailes}, {Kramer}, {Morello}, {Tabbara}, \& {van Straten}}]{Petroff16}
{Petroff}, E., {Barr}, E.~D., {Jameson}, A., {et~al.} 2016, \pasa, 33, e045

\bibitem[{{Petroff} {et~al.}(2019{\natexlab{b}}){Petroff}, {Oostrum},
  {Stappers}, {Bailes}, {Barr}, {Bates}, {Bhandari}, {Bhat}, {Burgay},
  {Burke-Spolaor}, {Cameron}, {Champion}, {Eatough}, {Flynn}, {Jameson},
  {Johnston}, {Keane}, {Keith}, {Kramer}, {Levin}, {Morello}, {Ng}, {Possenti},
  {Ravi}, {van Straten}, {Thornton}, \& {Tiburzi}}]{Petroff19}
{Petroff}, E., {Oostrum}, L.~C., {Stappers}, B.~W., {et~al.}
  2019{\natexlab{b}}, \mnras, 482, 3109

\bibitem[{{Piro}(2016)}]{Piro16}
{Piro}, A.~L. 2016, \apjl, 824, L32

\bibitem[{{Piro} \& {Burke-Spolaor}(2017)}]{Piro17}
{Piro}, A.~L., \& {Burke-Spolaor}, S. 2017, \apjl, 841, L30

\bibitem[{{Piro} \& {Gaensler}(2018)}]{Piro18}
{Piro}, A.~L., \& {Gaensler}, B.~M. 2018, \apj, 861, 150

\bibitem[{{Planck Collaboration} {et~al.}(2016){Planck Collaboration}, {Ade},
  {Aghanim}, {Arnaud}, {Ashdown}, {Aumont}, {Baccigalupi}, {Banday},
  {Barreiro}, {Bartlett}, {Bartolo}, {Battaner}, {Battye}, {Benabed},
  {Beno{\^\i}t}, {Benoit-L{\'e}vy}, {Bernard}, {Bersanelli}, {Bielewicz},
  {Bock}, {Bonaldi}, {Bonavera}, {Bond}, {Borrill}, {Bouchet}, {Boulanger},
  {Bucher}, {Burigana}, {Butler}, {Calabrese}, {Cardoso}, {Catalano},
  {Challinor}, {Chamballu}, {Chary}, {Chiang}, {Chluba}, {Christensen},
  {Church}, {Clements}, {Colombi}, {Colombo}, {Combet}, {Coulais}, {Crill},
  {Curto}, {Cuttaia}, {Danese}, {Davies}, {Davis}, {de Bernardis}, {de Rosa},
  {de Zotti}, {Delabrouille}, {D{\'e}sert}, {Di Valentino}, {Dickinson},
  {Diego}, {Dolag}, {Dole}, {Donzelli}, {Dor{\'e}}, {Douspis}, {Ducout},
  {Dunkley}, {Dupac}, {Efstathiou}, {Elsner}, {En{\ss}lin}, {Eriksen},
  {Farhang}, {Fergusson}, {Finelli}, {Forni}, {Frailis}, {Fraisse},
  {Franceschi}, {Frejsel}, {Galeotta}, {Galli}, {Ganga}, {Gauthier}, {Gerbino},
  {Ghosh}, {Giard}, {Giraud-H{\'e}raud}, {Giusarma}, {Gjerl{\o}w},
  {Gonz{\'a}lez-Nuevo}, {G{\'o}rski}, {Gratton}, {Gregorio}, {Gruppuso},
  {Gudmundsson}, {Hamann}, {Hansen}, {Hanson}, {Harrison}, {Helou},
  {Henrot-Versill{\'e}}, {Hern{\'a}ndez-Monteagudo}, {Herranz}, {Hildebrand t},
  {Hivon}, {Hobson}, {Holmes}, {Hornstrup}, {Hovest}, {Huang}, {Huffenberger},
  {Hurier}, {Jaffe}, {Jaffe}, {Jones}, {Juvela}, {Keih{\"a}nen}, {Keskitalo},
  {Kisner}, {Kneissl}, {Knoche}, {Knox}, {Kunz}, {Kurki-Suonio}, {Lagache},
  {L{\"a}hteenm{\"a}ki}, {Lamarre}, {Lasenby}, {Lattanzi}, {Lawrence}, {Leahy},
  {Leonardi}, {Lesgourgues}, {Levrier}, {Lewis}, {Liguori}, {Lilje},
  {Linden-V{\o}rnle}, {L{\'o}pez-Caniego}, {Lubin}, {Mac{\'\i}as-P{\'e}rez},
  {Maggio}, {Maino}, {Mandolesi}, {Mangilli}, {Marchini}, {Maris}, {Martin},
  {Martinelli}, {Mart{\'\i}nez-Gonz{\'a}lez}, {Masi}, {Matarrese}, {McGehee},
  {Meinhold}, {Melchiorri}, {Melin}, {Mendes}, {Mennella}, {Migliaccio},
  {Millea}, {Mitra}, {Miville-Desch{\^e}nes}, {Moneti}, {Montier}, {Morgante},
  {Mortlock}, {Moss}, {Munshi}, {Murphy}, {Naselsky}, {Nati}, {Natoli},
  {Netterfield}, {N{\o}rgaard-Nielsen}, {Noviello}, {Novikov}, {Novikov},
  {Oxborrow}, {Paci}, {Pagano}, {Pajot}, {Paladini}, {Paoletti}, {Partridge},
  {Pasian}, {Patanchon}, {Pearson}, {Perdereau}, {Perotto}, {Perrotta},
  {Pettorino}, {Piacentini}, {Piat}, {Pierpaoli}, {Pietrobon}, {Plaszczynski},
  {Pointecouteau}, {Polenta}, {Popa}, {Pratt}, {Pr{\'e}zeau}, {Prunet},
  {Puget}, {Rachen}, {Reach}, {Rebolo}, {Reinecke}, {Remazeilles}, {Renault},
  {Renzi}, {Ristorcelli}, {Rocha}, {Rosset}, {Rossetti}, {Roudier},
  {Rouill{\'e} d'Orfeuil}, {Rowan-Robinson}, {Rubi{\~n}o-Mart{\'\i}n},
  {Rusholme}, {Said}, {Salvatelli}, {Salvati}, {Sandri}, {Santos},
  {Savelainen}, {Savini}, {Scott}, {Seiffert}, {Serra}, {Shellard}, {Spencer},
  {Spinelli}, {Stolyarov}, {Stompor}, {Sudiwala}, {Sunyaev}, {Sutton},
  {Suur-Uski}, {Sygnet}, {Tauber}, {Terenzi}, {Toffolatti}, {Tomasi},
  {Tristram}, {Trombetti}, {Tucci}, {Tuovinen}, {T{\"u}rler}, {Umana},
  {Valenziano}, {Valiviita}, {Van Tent}, {Vielva}, {Villa}, {Wade}, {Wandelt},
  {Wehus}, {White}, {White}, {Wilkinson}, {Yvon}, {Zacchei}, \&
  {Zonca}}]{Planck16}
{Planck Collaboration}, {Ade}, P.~A.~R., {Aghanim}, N., {et~al.} 2016, \aap,
  594, A13

\bibitem[{{Platts} {et~al.}(2018){Platts}, {Weltman}, {Walters}, {Tendulkar},
  {Gordin}, \& {Kandhai}}]{Platts18}
{Platts}, E., {Weltman}, A., {Walters}, A., {et~al.} 2018, arXiv e-prints

\bibitem[{{Popov} \& {Postnov}(2010)}]{Popov10}
{Popov}, S.~B., \& {Postnov}, K.~A. 2010, in Evolution of Cosmic Objects
  through their Physical Activity, ed. H.~A. {Harutyunian}, A.~M. {Mickaelian},
  \& Y.~{Terzian}, 129--132

\bibitem[{{Prochaska} \& {Zheng}(2019)}]{Prochaska19}
{Prochaska}, J.~X., \& {Zheng}, Y. 2019, \mnras, 258

\bibitem[{{Ravi}(2019)}]{Ravi19}
{Ravi}, V. 2019, \mnras, 482, 1966

\bibitem[{{Ravi} {et~al.}(2016){Ravi}, {Shannon}, {Bailes}, {Bannister},
  {Bhandari}, {Bhat}, {Burke-Spolaor}, {Caleb}, {Flynn}, {Jameson}, {Johnston},
  {Keane}, {Kerr}, {Tiburzi}, {Tuntsov}, \& {Vedantham}}]{Ravi16}
{Ravi}, V., {Shannon}, R.~M., {Bailes}, M., {et~al.} 2016, Science, 354, 1249

\bibitem[{{Schmidt}(1968)}]{1968ApJ...151..393S}
{Schmidt}, M. 1968, \apj, 151, 393

\bibitem[{{Scholz} {et~al.}(2016){Scholz}, {Spitler}, {Hessels}, {Chatterjee},
  {Cordes}, {Kaspi}, {Wharton}, {Bassa}, {Bogdanov}, {Camilo}, {Crawford},
  {Deneva}, {van Leeuwen}, {Lynch}, {Madsen}, {McLaughlin}, {Mickaliger},
  {Parent}, {Patel}, {Ransom}, {Seymour}, {Stairs}, {Stappers}, \&
  {Tendulkar}}]{2016ApJ...833..177S}
{Scholz}, P., {Spitler}, L.~G., {Hessels}, J.~W.~T., {et~al.} 2016, \apj, 833,
  177

\bibitem[{{Scholz} {et~al.}(2017){Scholz}, {Bogdanov}, {Hessels}, {Lynch},
  {Spitler}, {Bassa}, {Bower}, {Burke-Spolaor}, {Butler}, {Chatterjee},
  {Cordes}, {Gourdji}, {Kaspi}, {Law}, {Marcote}, {McLaughlin}, {Michilli},
  {Paragi}, {Ransom}, {Seymour}, {Tendulkar}, \&
  {Wharton}}]{2017ApJ...846...80S}
{Scholz}, P., {Bogdanov}, S., {Hessels}, J.~W.~T., {et~al.} 2017, \apj, 846, 80

\bibitem[{{Shannon} {et~al.}(2018){Shannon}, {Macquart}, {Bannister}, {Ekers},
  {James}, {Os{\l}owski}, {Qiu}, {Sammons}, {Hotan}, {Voronkov}, {Beresford},
  {Brothers}, {Brown}, {Bunton}, {Chippendale}, {Haskins}, {Leach},
  {Marquarding}, {McConnell}, {Pilawa}, {Sadler}, {Troup}, {Tuthill},
  {Whiting}, {Allison}, {Anderson}, {Bell}, {Collier}, {G{\"u}rkan}, {Heald},
  \& {Riseley}}]{Shannon18}
{Shannon}, R.~M., {Macquart}, J.~P., {Bannister}, K.~W., {et~al.} 2018, \nat,
  562, 386

\bibitem[{{Spitler} {et~al.}(2016{\natexlab{a}}){Spitler}, {Scholz}, {Hessels},
  {Bogdanov}, {Brazier}, {Camilo}, {Chatterjee}, {Cordes}, {Crawford},
  {Deneva}, {Ferdman}, {Freire}, {Kaspi}, {Lazarus}, {Lynch}, {Madsen},
  {McLaughlin}, {Patel}, {Ransom}, {Seymour}, {Stairs}, {Stappers}, {van
  Leeuwen}, \& {Zhu}}]{Spitler16}
{Spitler}, L.~G., {Scholz}, P., {Hessels}, J.~W.~T., {et~al.}
  2016{\natexlab{a}}, \nat, 531, 202

\bibitem[{{Spitler} {et~al.}(2016{\natexlab{b}}){Spitler}, {Scholz}, {Hessels},
  {Bogdanov}, {Brazier}, {Camilo}, {Chatterjee}, {Cordes}, {Crawford},
  {Deneva}, {Ferdman}, {Freire}, {Kaspi}, {Lazarus}, {Lynch}, {Madsen},
  {McLaughlin}, {Patel}, {Ransom}, {Seymour}, {Stairs}, {Stappers}, {van
  Leeuwen}, \& {Zhu}}]{2016Natur.531..202S}
---. 2016{\natexlab{b}}, \nat, 531, 202

\bibitem[{{Tendulkar} {et~al.}(2017){Tendulkar}, {Bassa}, {Cordes}, {Bower},
  {Law}, {Chatterjee}, {Adams}, {Bogdanov}, {Burke-Spolaor}, {Butler},
  {Demorest}, {Hessels}, {Kaspi}, {Lazio}, {Maddox}, {Marcote}, {McLaughlin},
  {Paragi}, {Ransom}, {Scholz}, {Seymour}, {Spitler}, {van Langevelde}, \&
  {Wharton}}]{Tendulkar17}
{Tendulkar}, S.~P., {Bassa}, C.~G., {Cordes}, J.~M., {et~al.} 2017, \apj, 834,
  L7

\bibitem[{{Thornton} {et~al.}(2013){Thornton}, {Stappers}, {Bailes},
  {Barsdell}, {Bates}, {Bhat}, {Burgay}, {Burke-Spolaor}, {Champion}, {Coster},
  {D'Amico}, {Jameson}, {Johnston}, {Keith}, {Kramer}, {Levin}, {Milia}, {Ng},
  {Possenti}, \& {van Straten}}]{Thornton13}
{Thornton}, D., {Stappers}, B., {Bailes}, M., {et~al.} 2013, Science, 341, 53

\bibitem[{{Totani}(2013)}]{Totani13}
{Totani}, T. 2013, Publications of the Astronomical Society of Japan, 65, L12

\bibitem[{{Turolla} {et~al.}(2015){Turolla}, {Zane}, \& {Watts}}]{Turolla15}
{Turolla}, R., {Zane}, S., \& {Watts}, A.~L. 2015, Reports on Progress in
  Physics, 78, 116901

\bibitem[{{Wang} \& {Yu}(2017)}]{2017JCAP...03..023W}
{Wang}, F.~Y., \& {Yu}, H. 2017, Journal of Cosmology and Astroparticle
  Physics, 3, 023

\bibitem[{{Wang} {et~al.}(2016){Wang}, {Yang}, {Wu}, {Dai}, \& {Wang}}]{Wang16}
{Wang}, J.-S., {Yang}, Y.-P., {Wu}, X.-F., {Dai}, Z.-G., \& {Wang}, F.-Y. 2016,
  \apj, 822, L7

\bibitem[{{Zhang}(2014)}]{Zhang14}
{Zhang}, B. 2014, \apj, 780, L21

\bibitem[{{Zhang}(2018)}]{Zhang18}
---. 2018, \apjl, 867, L21

\end{thebibliography}

\end{document}